\documentclass{article}
\usepackage{authblk}
\usepackage[a4paper, left=3cm, right=3cm, top=2.0cm, bottom=2.0cm]{geometry}
\usepackage{mathptmx}   
\usepackage{helvet}          
\usepackage{courier}      
\usepackage{tabularx}
\newcolumntype{C}{>{\centering\arraybackslash}X}
\usepackage{caption}
\usepackage{amsmath}
\usepackage{amssymb}
\usepackage{textcomp}
\usepackage{amsfonts}
\usepackage{float}
\usepackage{wrapfig}
\usepackage{sidecap}
\usepackage{graphicx}
\usepackage{helvet}
\usepackage{color}
\usepackage[authoryear]{natbib}
\usepackage{textcomp}
\usepackage{siunitx}
\usepackage{upgreek}
\usepackage{sidecap}
\usepackage{lineno}

\usepackage{array} 
\usepackage{IEEEtrantools}
\usepackage{graphicx}
\usepackage{float}
\usepackage{tabularx}
\usepackage{rotating}
\usepackage{upgreek}
\usepackage{amsmath}
\usepackage{amssymb}
\usepackage{amsfonts}

\usepackage{xcolor}
\newcommand{\mycirc}[1][black]{\normalsize\textcolor{#1}{\ensuremath\bullet}}

\usepackage{microtype}
\usepackage{textcomp}
\usepackage{setspace}
\usepackage[latin1]{inputenc}
\nopagebreak
\title{\vspace{-2.2cm}Nonlinear response properties in the inferior colliculus: Common sound-frequency dependence for linear and nonlinear responses}
\author[1]{Dominika Lyzwa}%\footnote{Corresponding author: dlyzwa@gwdg.de}}
\author[2,3]{Chen Chen}
\author[2,3]{Monty A. Escab\'{i}}
\author[3,4]{Heather L. Read}
\affil[1]{Max Planck Institute for Dynamics and Self-Organization, G\"ottingen, Germany}
\affil[2]{Department of Electrical Engineering, University of Connecticut, USA}
\affil[3]{Department of Biomedical Engineering, University of Connecticut, USA}
\affil[4]{Department of Psychological Sciences, University of Connecticut, USA}	
\date{}
\begin{document}
\maketitle
\vspace{-0.5cm}
Neurons in %the central nucleus of 
the main center of convergence in the auditory midbrain, the central nucleus of the inferior colliculus (ICC) have been shown to display either linear significant receptive fields, or both, linear and nonlinear significant receptive fields. In this study, we used reverse correlation to probe linear and nonlinear response properties of single neurons in the cat ICC. The receptive fields display areas of stimulus parameters leading to enhanced or inhibited spiking activity, and thus allow investigating the interplay to process complex sounds. Spiking responses were obtained from neural recordings of anesthetized cats in response to dynamic moving ripple (DMR) stimuli. The DMR sound contains amplitude and frequency modulations and allows systematically mapping neural preferences. Correlations of the stimulus envelope that preferably excite neurons can be mapped with the spike-triggered covariance. The spike-triggered average and -covariance were computed for the envelope of the DMR, separately for each frequency carrier (spanning a range of 0-5.5 octaves). This enables studying processing of the sound envelope, and to investigate whether nonlinearities are more pronounced at the neurons' preferred frequencies rather than at other frequencies. We find that more than half of the neurons (n=120) display significant nonlinear response properties at least at one frequency carrier. Nonlinearities are dominant at the neuron's best frequency. The nonlinear preferences can have either the same or opposite temporal receptive field pattern (e.g. on-off) as the linear preferences. No relationship to other neural properties such as feature-selectivity, phase-locking, or the like has been found. Thus, these nonlinearities do not seem to be linked to a specific type of neuron but to be inherent to ICC neurons indicating a diverse range of filtering characteristics. 
\section*{Introduction}% 
Vocalizations can display amplitude and frequency modulations, harmonics and envelope correlations, as well as temporal and spectral correlations. %Eingefuegt:
We investigated the encoding of vocalizations by clusters of auditory midbrain neurons and demonstrated that within each neuronal cluster, there is responsiveness to almost all vocalizations \citep{LyzwaClass2016a, LyzwaClass2016b}. %
How does this responsiveness to vocalizations manifest at the single neuron level? Do preferences exist to specific acoustic features of the vocalizations?\\ %
Neurons in the auditory midbrain can employ various processing mechanisms --linear and nonlinear ones-- in order to encode these behaviorally relevant sounds. %
A neuron whose response to complex sound can be described as a linear superposition of the responses to its (independent) constituents has linear response properties, and a linear response is proportional to the input (e.g.~increased spiking rate for enhanced stimulus intensity). A neuron which does not display these characteristics has nonlinear response properties. %
In the inferior colliculus (IC), an important part of the neurons % 
display nonlinear response properties \citep{Monty2002}.\\ %
For the visual system it has been suggested that simple receptive fields which display linear summation properties form a unit with a surrounding complex receptive field that displays nonlinear summation properties \citep{HubelWiesel1963, AllmanMcGuinness1985a, AllmanMcGuinness1985b}, and that this might increase selectivity of neurons \citep{GallantRF2002}.\\ %
%\newpage
In this work, linear and nonlinear auditory receptive fields are investigated. Do the nonlinear responses lie within the linear spectrotemporal receptive field? Or do such nonlinearities lie outside, thus forming a non-classical receptive field without nonlinear summation analogous to cells in the visual V1 cortex \citep{GallantRF2002}. %
Since natural sounds are composed of several temporal and spectral correlations, they do not allow directly linking the neural response to a particular stimulus property, but only to a composition of a spectral and temporal component. A systematic approach to identify neuronal preferences, on the other hand, is offered by the use of the dynamic moving ripple sound (DMR) which contains dynamically varying combinations of frequency and amplitude modulations. This artificial sound can be parametrically manipulated, allowing for the characterization of neuronal preferences over a broad spectrotemporal range \citep{Monty2002}.\\ % 
Neurons were shown to display preferences, enhanced spiking, for certain amplitude and frequency modulations \citep{WoolleyTuningDisc2005, Monty2002}, and it was suggested that such preferences for specific modulations of individual vocalizations facilitate encoding of these behaviorally relevant sounds \citep{WoolleyTuningDisc2005}. Neurons might also display preferences to envelope correlations. Preferences to specific stimulus correlations would be due to nonlinear response mechanisms. In the ICC, \(\sim40\%\) of the neurons were shown to display nonlinear response properties \citep{Monty2002}, and different nonlinear response types exist \citep{Monty2002, McAlpine2004, MontySpkThres2005, AndoniPollakFMselecSTC2011}.\\ 
Here, we investigated neural preferences to stimulus correlations which are also present in natural sounds, at the single cell level. In order to assess in which frequency range the nonlinearities are dominant, correlation preferences were analyzed separately for each frequency carrier of the presented sound. Linear and nonlinear response properties of neurons are visualized with receptive fields that can capture those preferences in response to a specific stimulus. Comparison of linear and nonlinear response properties can yield insight into whether the nonlinearity is due to intrinsic processing, in which case the receptive field would be similar, or, if it is already present in the input of the neuron, in which case the receptive field would be disparate.
Using first and second order reverse correlation, the linear and nonlinear receptive fields are computed and compared with respect to spectrotemporal preferences.\\
The reverse correlation method is well suited for the derivation of nonlinear stimulus preferences since it does not put any limit on the number of extracted stimulus features, as opposed to the also often used maximally informative dimension method \citep{SharpeeMID2004,SharpeeCooperative2008, SharpeeReviewHierarchAudio2011}. Furthermore, the dynamic moving ripple stimulus meets the requirements for use with the reverse correlation method. %
The modulation spectrum of the DMR is used for the reverse correlation analysis. Using the envelope of the DMR sound allows studying the representation of amplitude modulations in the ICC. Amplitude modulations are essential for recognition of complex sounds such as speech \citep{ShannonTemp1995} and vocalizations. % 
At the level of the ICC, frequencies (\(>\)~1~kHz) are not encoded due to the inability of the neurons to follow the fast changes \citep{Palmer2006}. The frequency carriers, the fine structure, which were used for the DMR sound span a frequency range of 1-48~kHz.\\
In previous studies, the entire auditory sound waveform was considered for the reverse correlation analysis \citep{DijkWiener1994, YamadaLewis1999} or for derived information theoretic approaches \citep{AndoniPollakFMselecSTC2011}. In the present work however, the reverse correlation method is applied to the spectrotemporal envelope, individually for each frequency carrier. 
Nonlinear response properties to the envelope have been observed in several processing stations of the auditory ascending pathways and become increasingly complex and specialized as one progresses up the auditory hierarchy \citep{SahaniLinden2003, SharpeeCooperative2008, TheunissenSTRF2000}. However, it has not been investigated previously in which frequency range these nonlinearities are contained. By computing the reverse correlation for each frequency carrier of the DMR sound, it is possible to assess in which frequency range the nonlinearities are dominant and how these frequencies relate to the best frequency of the neuron.\\
\section*{Material \& Methods}%

\subsubsection*{Electrophysiology}%
Recordings were made from the ICC of anesthetized cats, %
in response to dynamic moving ripple sound (DMR) sound, and isolation of single neuron responses was performed.\\ 
From five adult female cats recordings from the %right?
left ICC were taken. Following surgery, described in \citep{Monty2002, EscabiReadEfficient2010, ChenSparse2012}, %ChenNeighborSame2012}  
the animal was kept anesthetized in a nonreflexive state by continuous infusion of Ketamine and Diazepam. Throughout the experiment, heart and breathing rate, temperature and reflexes were monitored and the infusion rate was adjusted accordingly. Neural recordings were performed over a period of 24-72~h. The probes were first positioned on the surface of the IC with a stereotaxic frame (Kopf Instruments) at an angle of 30\(^\circ\) relative to the sagittal plane, orthogonal to the frequency-band lamina \citep{LangnerLaminar1997}, and successively inserted deeper into the ICC.\ 
Acute 4-tetrode (16~channel) recording probes (two shanks with two tetrode sites on each, 150~\( \upmu \)m spacing, impedance 1.5-3.5~M\( \Omega\) at 1~kHz, NeuroNexus Technologies) were used to record neuronal activity from the ICC. Best frequencies span a range of 1.3-16.7~kHz. %
The experimental setup is described in detail in \citep{ChenNeighborSame2012,ChenSparse2012}.%and the DMR stimulus are, Monty2002
\subsubsection*{Dynamic moving ripple sound stimulus (DMR)}%
The dynamic moving ripple (DMR) sound stimulus was presented acoustically to the cats, dichotically, in reverse order. At a fixed intensity of 80~dB~SPL, a 10~min sequence of the DMR was played twice (20 min in total),  % 
yielding two trials, A and B, of the recording. %
\\
The dynamic moving ripple stimulus sound contains amplitude and frequency modulations which are present in natural sounds. 
Compared with natural sound, the synthetic DMR sound offers the advantage that it can be %
parametrically manipulated, %
thus allowing for a systematic characterization of neuronal spectrotemporal preferences. %
The DMR sound has been widely used \citep{SharpeeCooperative2008, EscabiTradeoff2010}. This sound stimulus is described in detail in \citep{Monty2002}. %
\\
In order to create the DMR sound, energy modulation in time and in frequency is applied to a bank of sinusoidal (\(n=\)~659) carriers of frequency \(f_{k}\). The frequency carriers span a range between 0-5.8~octaves, with a spectral resolution of 0.0085~octaves. The ripple density \( \Omega (t)\) defines the number of spectral peaks per octave at a time instant, and \( F_{\mathrm{m}}(t)\) defines the instantaneous temporal modulation. These two values are independent and vary slowly in time: \(\Omega (t)\) varies within 0-4~cycles/octave with a maximum rate of change of 3~Hz, and \( F_{\mathrm{m}}(t)\) varies between 0-500~Hz at a maximum rate of 1.5~Hz. % 
Within these parameter ranges amplitudes are uniformly distributed in order to cover the stimulus space in a statistically unbiased manner. %
The rates of change have been taken from observed rates for similar features in speech and vocalizations \citep{Greenberg1999, Monty2002}. The parameters vary at a slow rate compared with the integration time of ICC neurons, which is about \(\sim\)10~ms \citep{ChenSparse2012}. The peak-to-peak amplitude of the DMR was set to 30~dB since this produces robust responses for the vast majority of IC neurons \citep{ContrastEscabi2003}. %
\\ 
The formula of the DMR sound envelope is given by %
 \(S_{\mathrm{\textsf{DMR}}}(X_{k}, t) = M/2\cdot \mathrm{sin}\left(2\pi \Omega(t)X_{\mathrm{k}} + \Phi_{\mathrm{k}}\right)\)%
%\end{eqnarray}%
, where \(M\) is the modulation depth (\(M\)=30), % the ripple density \(\Omega (t)\), 
\(\Phi_{\mathrm{k}}\) is the integrated modulation rate of \(F_{\mathrm{m}}(t)\) and \(\Omega(t)\), and \(X_{\mathrm{k}}\) is the frequency axis in octaves relative to the lowest stimulus frequency (\(f_{\mathrm{0}} = 1000\)~Hz). The acoustic waveform which was played to the animals is obtained by adding all modulated sinusoidal frequency carriers %.
with a randomly chosen phase %
between 0-2\(\pi\), which creates a noise-like character that is necessary for the reverse correlation analysis, and a modulating % 
transformed form of the %
spectrotemporal envelope. %
\\
The reverse correlation method % 
%requires 
assumes %
the input to be white noise \citep{LeeSchetzen1965}. However, it has been shown that neurons in higher auditory areas in primates are not well driven by white noise or simple stimuli \citep{Rauschecker1995}. % 
The DMR fulfills %
requirements for use with the reverse correlation methods, while containing the frequency and amplitude modulations and coherent modulations in the range which is characteristic for vocalizations \citep{Monty2002, TheunissenSTRF2000}. %
The local envelope statistics of the DMR are dynamic, they continuously vary with time.\newpage %
Short-term correlations exist and have been evaluated for several envelope segment durations \citep{Monty2002}. The global (long-term) autocorrelation consists of an impulse-like central peak of width 3~ms and a quarter of an octave \citep{Monty2002}. A narrow impulse-like character of the autocorrelation is a prerequisite to use reverse correlation analysis for the derivation of receptive fields \citep{Eggermont1993, TheunissenSTRF2000}. % 
The DMR allows probing neural response preferences with the reverse correlation method. %
\subsubsection*{Data processing}%
Electrophysiological recordings were processed to yield single neuron responses. %\\
Neural responses were recorded with a RX5 Pentusa Base station (Tucker-Davis Technologies) followed by offline analysis in MATLAB (MathWorks Inc.). The continuous neural traces were digitally bandpass filtered (300-5000~Hz) and were used to detect spikes exceeding 5 standard deviations of the spontaneous activity. The candidate action potentials and spike waveforms were aligned and sorted using peak values and first principle components with an automated clustering KlustaKwik software \citep{HarrisAccuracy2000}. The single neurons responses were used for the reverse correlation analysis. 
\subsubsection*{Reverse correlation: Spike-triggered average and covariance}%
Receptive fields can show e.g. the averaged spiking activity in response to stimulus frequencies, the neuron's corresponding delay and integration time (spectrotemporal receptive field) or the averaged spiking activity to stimulus correlations (spike-triggered covariance).%
\\
Neural ICC responses were analyzed with respect to the second order Wiener kernel in order to test whether they display preferences to these nonlinear stimulus interactions. 
The STA and STC receptive fields are derived from the kernels of the Wiener expansion \citep{LeeSchetzen1965}, an expansion of functionals which approximates nonlinear systems \citep{MarmarelisNaka1972, IdentiMultiInputMarmarelis1974, Aertsen1981a}. % 
It was shown that the Wiener kernels up to the second order (Eq.~\ref{eq:WienerKernelTerms}) can be described by %
the following expressions (Eq.~\ref{eq:WienerKernelCross}), adapted from \citep{DijkWiener1994}:
\begin{IEEEeqnarray}{crl}
r(t)  & = &  k_{0}+k_{1}+k_{2}
\label{eq:WienerKernelTerms}
\\
& = & \left\langle y(t) \right\rangle +\frac{1}{A}\left\langle y(t)\;x(t-\tau)\right\rangle
+ \frac{1}{2!A^{2}}\left\langle \left[ y(t)-k_{0} \right]\;x(t-\tau_{1})\;x(t-\tau_{2})\right\rangle  
\label{eq:WienerKernelCross}  
\\
& \cong &  \frac{1}{T}{\int}^{T}_{0} dt \sum^{N}_{i = 1} \delta (t-t_{i})  
+ \frac{1}{AT}\int_0^T dt\; x(t-\tau_{1}) \sum^{N}_{i = 1}\delta (t-t_{i}) +\frac{N_{0}}{AN} \sum^{N}_{i=1}x(t_{i}-\tau) \nonumber\\
&& +\: \frac{1}{2A^{2}T} \int^T_0 \left( \sum^{N}_{i = 1}  \delta (t-t_{i}) -N_{0} \right)\times \ x(t-\tau_{1})\;x(t-\tau_{2})dt 
\label{eq:WienerKernelInt}
\\
& = & N_{0}+ \frac{N_{0}}{A N}\sum^{N}_{i=1}x(t_{i}-\tau)\nonumber\\
&& +\: \frac{N_{0}}{2!A^{2}}\left[\frac{1}{N}\sum^{N}_{i=1}x(t-\tau_{1})\;x(t-\tau_{2})-\frac{1}{T} \int_0^T x(t)\; x(t-\tau)dt\right]
\label{eq:WienerKernelFinal}
\end{IEEEeqnarray}
where \(N\) is the total number of spikes, \(T\) is the entire recording time, \(A\) is the power spectral density of the stimulus, \(x(t)\) is the %
time-dependent input %
to the system, the stimulus, \(t_{i=1...N}\) are spiking times, \(y(t)\) is the temporal output of the system, the measured spike train, \(N_{0}=\) N/T is the average spike rate and \(\tau,~\tau_{1}\) and \(\tau_{2}\) are time delays from the interval \([0,T]\). Spike occurrences at times \(t_{i}\) are represented by \(\delta\)-function pulses, \(\delta(t-t_{i})\), and the entire spike train is the sum of all \(\delta\)-pulses. %
\\
The first term is the average system's response. The second term is obtained by cross-correlating the stimulus \(x(t)\) and the response \(y(t)\); the third term is obtained via a second-order cross-correlation of the stimulus and spike-train (Eq.~\ref{eq:WienerKernelCross}).%
From this expression, the terms in Eq.~\ref{eq:WienerKernelInt} are approximated due to finite recording times. In the final expression (Eq.~\ref{eq:WienerKernelFinal}), the first term represents the average spike rate, the second term the spike-triggered average, and the third term represents the spike-triggered covariance %.
which captures correlations within the stimulus that trigger the neuron to spike \citep{Schetzen1981,EggermontReview1983}. %
\subsubsection*{Spectro-temporal receptive field}%
The spike-triggered average (STA) is obtained by averaging the temporal windows of the DMR sound envelope that elicited a spike \citep{Monty2002}.
The STA for all frequency carriers \(f_{k}\) corresponds to the linear spectrotemporal receptive field (STRF). %
The STRF is the best linear model that transforms any time-varying stimulus into a prediction of the firing rate of a neuron \citep{Aertsen1981a, EggermontReview1983}. %
The STA in response to the envelope of the DMR sound stimulus is given by:
\begin{eqnarray}\label{eq:STRF}
\mathrm{\textsf{STA}(f_{k})} =  \frac{1}{\sigma^2_{\mathrm{DMR}}T}\sum^{N}_{i=1}{S_{\mathrm{DMR}}(f_{k}, t_{i}-\tau)}
\end{eqnarray} 
with the normalizing variance of the envelope \(\mathit{S_{\mathrm{DMR}}}\)% 
, \(\mathit{\sigma^2_{\mathrm{DMR}}=\mathrm{M^2/8}}\), recording time (\(\mathit{T~=}\)~600~s), and the delay \(\mathit{\tau}\) relative to the spike timing \(\mathit{t_i}\). This delay determines the temporal width of the window. %
\\
This receptive field is 2-dimensional, here, with a maximum time delay of 50~ms, of temporal resolution of 0.5~ms and a frequency range of 0-5.8~octaves, with 659 frequency channels \(f_{k}\), thus a fine spectral resolution of 0.0085~octaves. The maximum temporal delay of 50~ms was chosen because integration and delay times of ICC neurons are enclosed within this temporal window \citep{Monty2002, LangnerLatency1987}. %
\\
The statistically significant portion of the STRF is obtained by keeping all values of the STRF that exceed {3 \(\sigma\)~(1.6 \(\sigma\))} of the control noise STRF (which was obtained by adding random sound waveform segments) and setting all other values to zero \citep{Monty2002}. The STRFs have excitatory and/or inhibitory regions, which indicate respectively enhanced and suppressed spiking activity for these specific temporal and spectral parameters. % 
\\
For the STRF and for the STC-derivation, the full 20~min of recorded spiking activity and DMR sound are used. The STRF is well suited to map linear response properties \citep{Aertsen1981a, Aertsen1981b, TheunissenSTRF2000, KleinRipple2000}. However, to describe nonlinear response properties, the second order reverse correlation method, the spike-triggered covariance is necessary.
\subsubsection*{Spike-triggered covariance (STC)}%
\begin{figure}[h!t]\centering
   \includegraphics[width=0.45\linewidth]{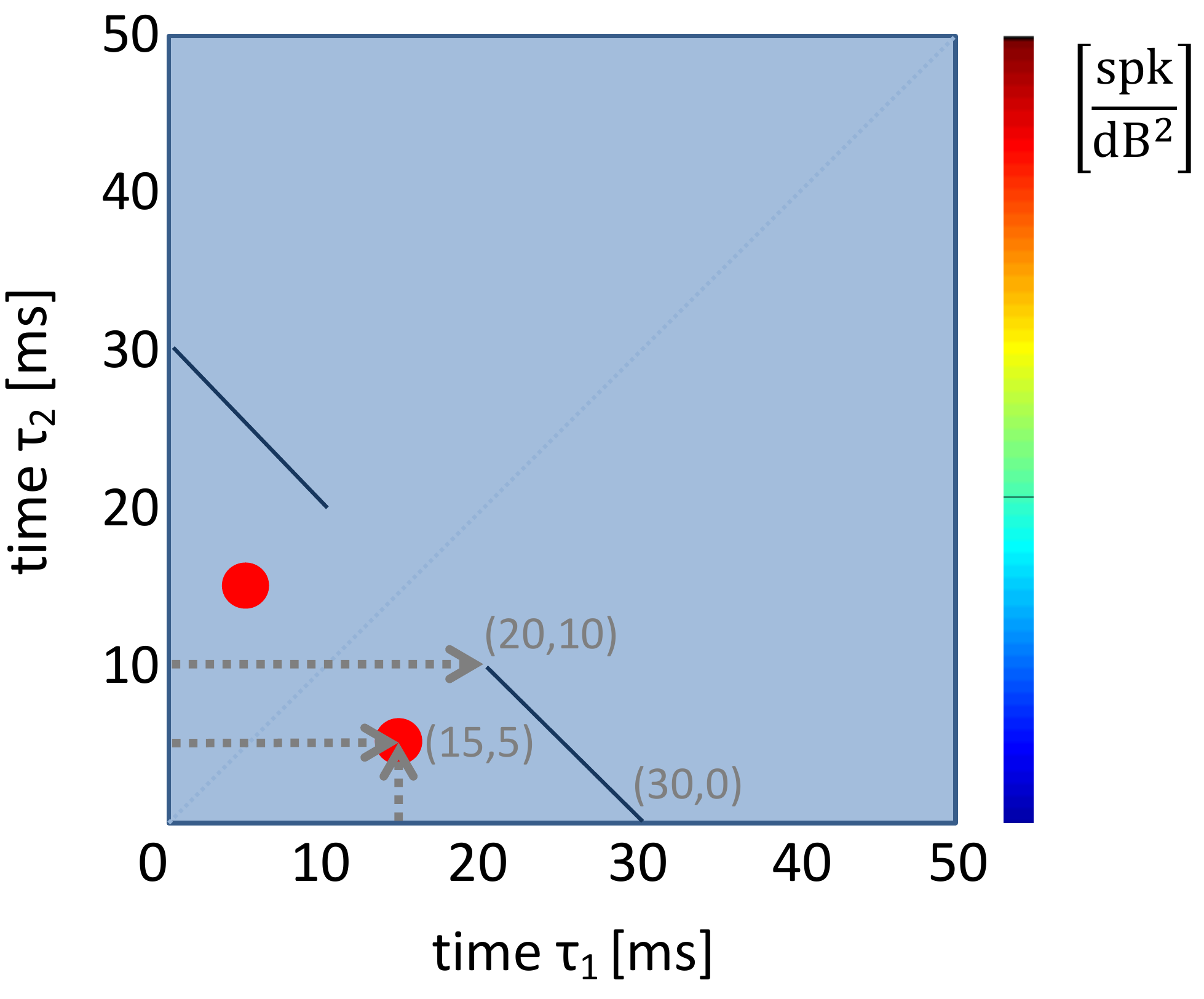}%
  \caption[Schematic spike-triggered covariance.]{Schematic example of spike-triggered covariance displaying excitatory (red) and inhibitory (blue) stimulus correlations. %
The matrix is symmetric around the diagonal, and therefore, the additional half-matrix does not provide further information, yet for better visualization the full matrix is displayed.}
\label{fig:exampleSTC}
\end{figure}

\begin{figure}[h!bp]\centering%
 \begin{minipage}[t]{0.31\textwidth}
   \includegraphics[width=0.95\textwidth]{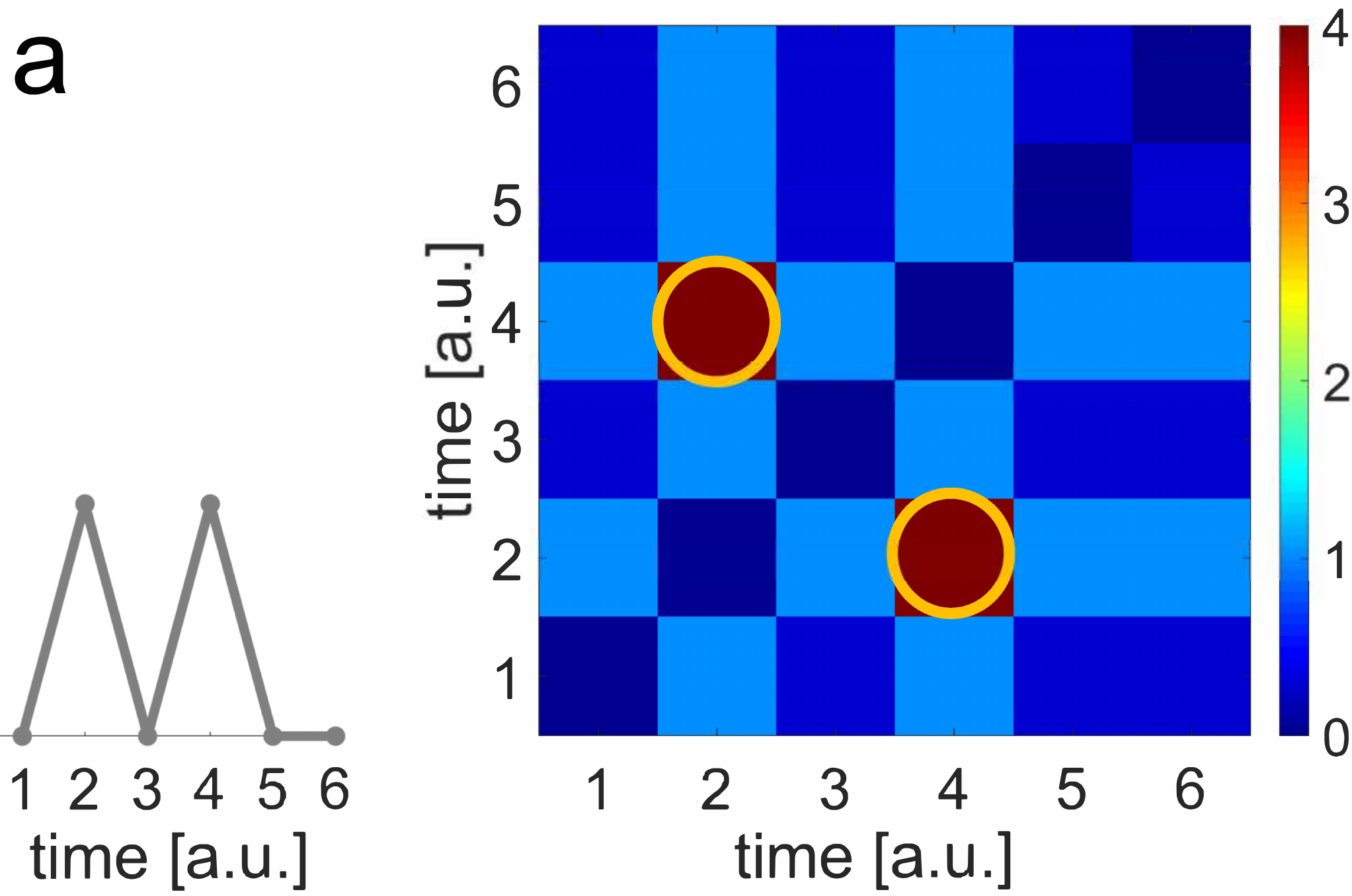}%
 \end{minipage}
\hspace{0.01\textwidth}
 \begin{minipage}[t]{0.30\textwidth}
   \includegraphics[width=\textwidth]{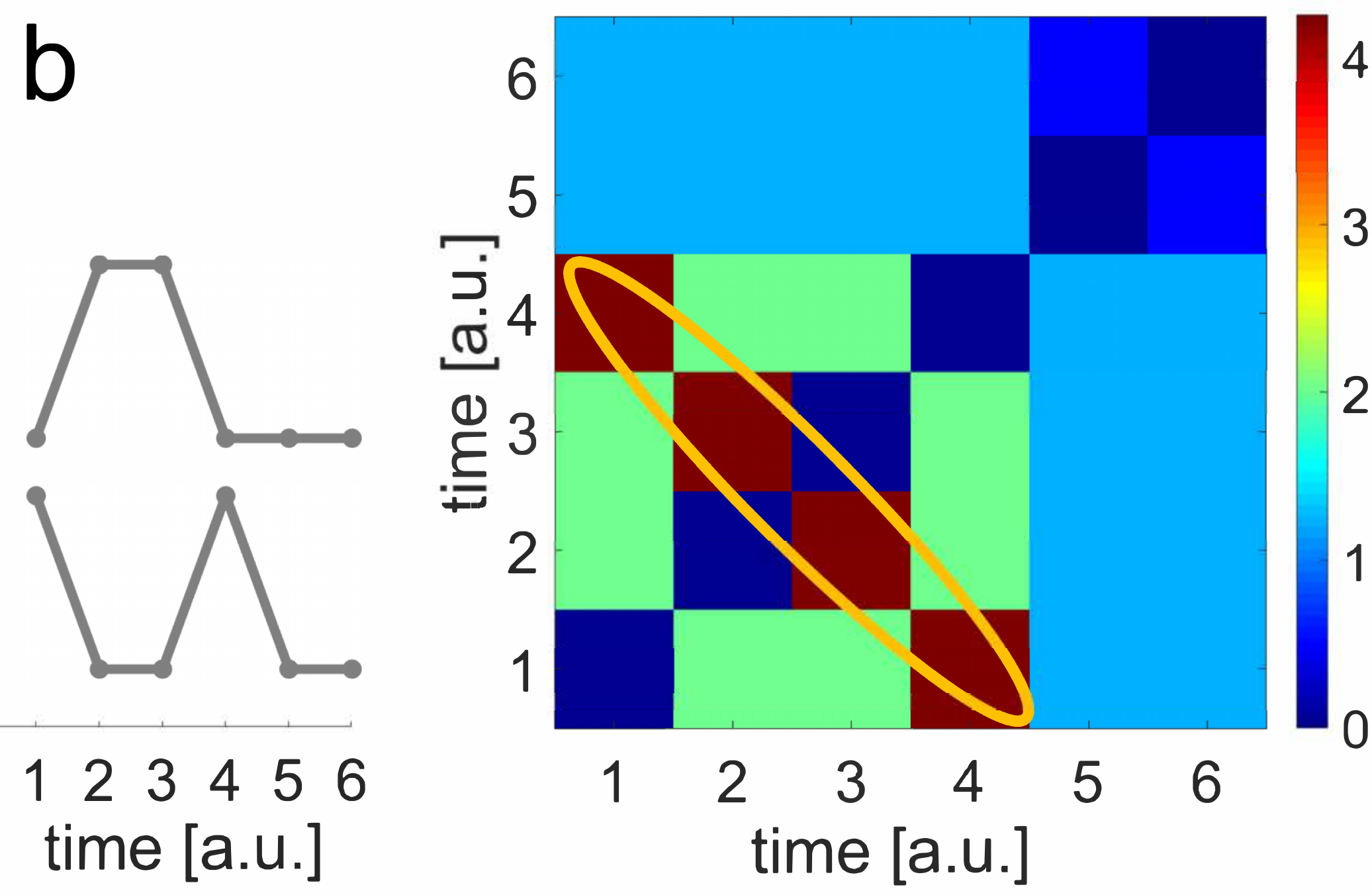}%
 \end{minipage}
\hspace{0.03\textwidth}
  \begin{minipage}[t]{0.30\textwidth}
   \includegraphics[width=\textwidth]{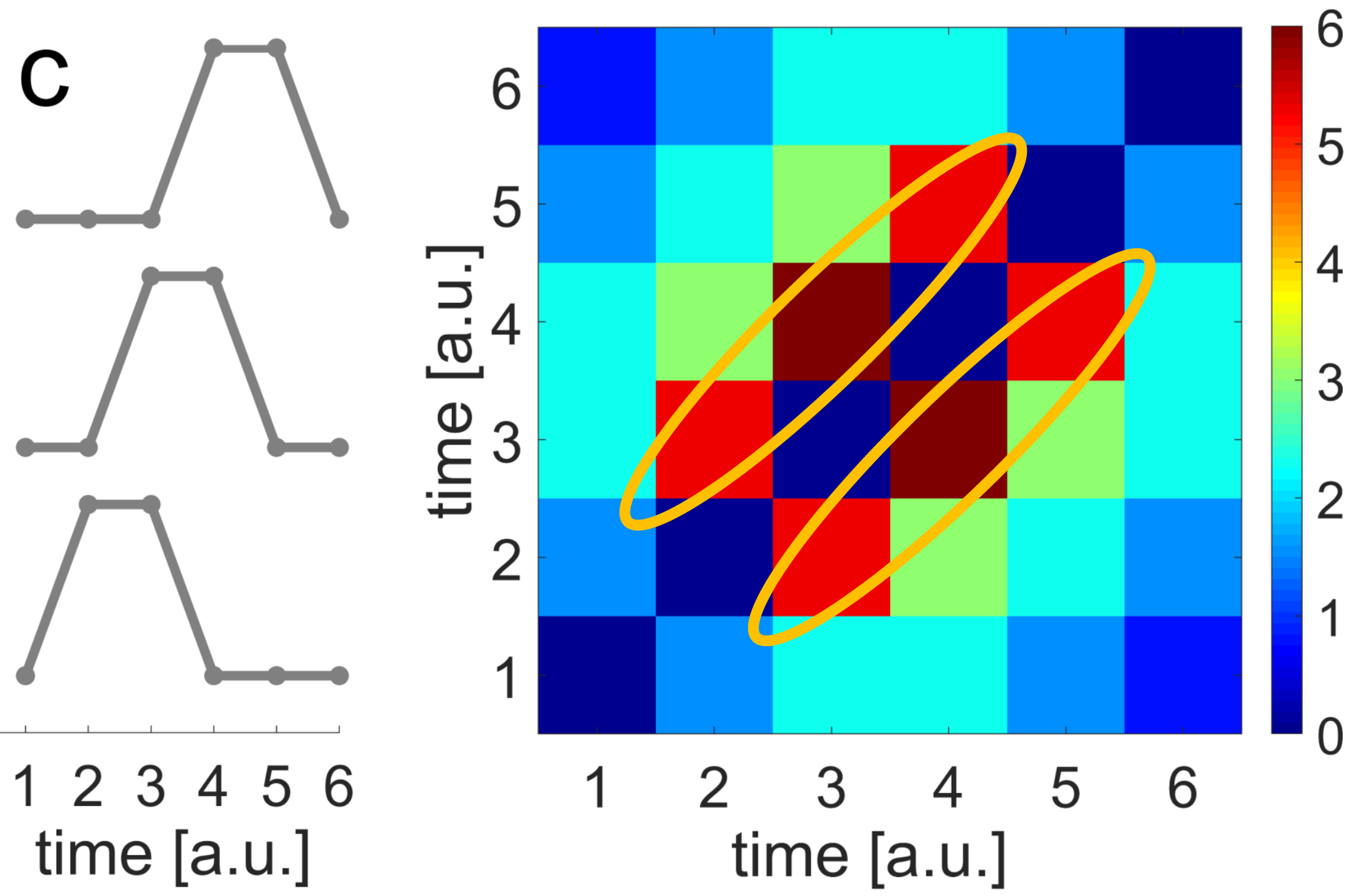}%
 \end{minipage}%
 \vfill
  \caption[Relating spike-eliciting stimulus correlations to STC shape.]{Relation of stimulus correlations and STC shape. Three examples of preferences to different stimulus correlations (left) and the derived outer products (right) are given a),~b) and c). Diagonals which correspond to zero delay have been subtracted in the outer products. Time units are arbitrary.}%
\label{fig:schematic}
\end{figure}

Nonlinear neural response preferences can be investigated with the spike-triggered covariance (STC). % here?
The expression for the spike-triggered covariance, derived in Eq.~\ref{eq:WienerKernelFinal}, specifically in response to the DMR sound envelope and separately for each frequency channel \(f\) is given by: %
\begin{eqnarray}\label{eq:WienerExpansion}
	\textsf{STC(f)} = \frac{N_{0}}{2!A^{2}}\left[\frac{1}{N}\sum^{N}_{i=1}S_{\mathrm{DMR}}(f, t_{i}-\tau_{1})S_{\mathrm{DMR}}(f, t_{i}-\tau_{2})-\frac{1}{T} \int^T_0 S_{\mathrm{DMR}}(f, t)S_{\mathrm{DMR}}(f, t-\tau)dt\right]
\end{eqnarray}
The first covariance is computed from all stimulus segments preceding a spike, the second part is the covariance of the stimulus. %
\\
The STC is obtained by comparing the variance of the entire stimulus distribution to the spike-eliciting variance. The STC describes the average correlations within the stimulus that lead to an elicitation of a spike. %
\\
While the average spike rate and spectrotemporal receptive field are rather intuitive, the spike-triggered covariance is not. %
In Figure~\ref{fig:exampleSTC}, a schematic example for the spike-triggered covariance is shown. %
The schematic displays an enhanced spike rate for stimulus correlations at time delays \(\mathit{(\tau_{1}, \tau_{2})} =\)~(15,5)~ms and~(5,15)~ms, thus at a relative delay of 10~ms. The spike rate is suppressed for stimulus correlations between time delays %
\(\mathit{(\tau_{1},\tau_{2})} =\)~(20,10)~ms and (30,0)~ms, respectively,
\(\mathit{(\tau_{1},\tau_{2})} =\)~(0,30)~ms and (10,20)~ms, thus a maximum relative delay of 10~ms. % 
In order to illustrate the relation of the preferred stimulus correlations and the shape of the spike-triggered covariance, a schematic with three examples is given in Fig.~\ref{fig:schematic}. %
\newpage
Correlations within the stimulus exist at \(\mathit{(\tau_{1}, \tau_{2}}) =\)~(2,4) and (4,2), and enhanced energy is present at these delay times in the outer product, which is derived from the stimulus (Fig.~\ref{fig:schematic}a). In the second example (Fig.~\ref{fig:schematic}b), stimulus correlations are present at \(\mathit{(\tau_{1},\tau_{2})} =\)~(1,4) and (4,1), and at \(\mathit{(\tau_{1},\tau_{2})} =\)~(2,3) and (3,2), with relative delays of 3 and 1~bins. These preferences give rise to an enhanced energy along the orthodiagonal. In the third example (Fig.~\ref{fig:schematic}c), three different stimulus correlations which occur at different times \(\mathit{\tau}\) but all have the same relative delay of 1~bin are present. These stimulus correlations give rise to enhanced energy along the paradiagonals, which are 1~bin away from the diagonal. For a larger relative delay, e.g.~5~bins, the paradiagonals will be 5~bins away from the diagonal. %
These shapes in the third example (at a distance of 0.5-1.5~ms from the diagonal) were found previously in the auditory nerve when computing the spike-triggered covariance in response to white noise \citep{LewisYamada2002}. %
We interpret the pattern of paradiagonals to represent a neuronal response component that is phase-locked to the square of the envelope of a filtered version of the stimulus waveform, or resulting from a quadrature pair of eigenvectors. %
\\
Unlike prior work in the auditory nerve, the STC derived in this work is obtained from the DMR sound envelope and calculated separately for each frequency channel \(f_{\mathrm{c}}\). This approach allows identifying dominant nonlinearities across frequency channels. %carriers.
\subsubsection*{Derivation STC}%
\begin{figure}[h!t]\centering
\includegraphics[width=0.8\linewidth]{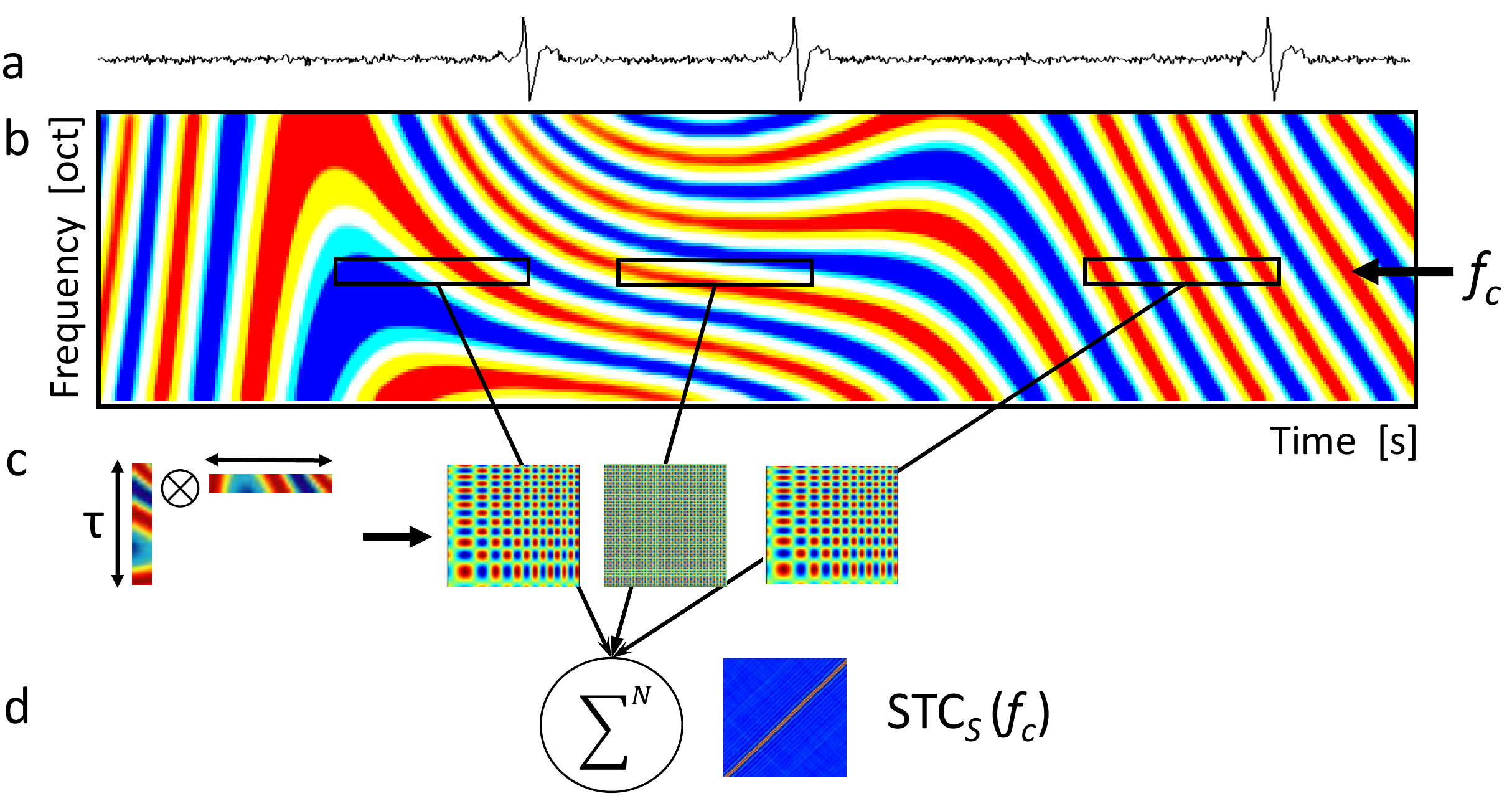}%
\caption[Schematic derivation of spike-triggered covariance.]{Schematic derivation of spike-triggered correlation matrix \(\mathrm{STC_{\mathrm{s}}}\). %
a) The top shows the first 500~ms of a single spike train obtained from recording while the DMR sound with the amplitude modulation spectrum shown in b) was presented. When a spike occurred, an immediately preceding time window of \(\mathit{\tau}\!=\)50~ms was taken from the spectrum for a single frequency carrier \(\mathit{f}\). c) From this vector the outer product was computed% . %
, respectively for each segment preceding a spike. %
d) The spike-triggered correlation matrix for \(\mathit{f}\) is the average of all N obtained outer products.% 
}\label{fig:generateSTC}
\end{figure}

\begin{figure}[th!]\centering
   \includegraphics[width=0.99\linewidth]{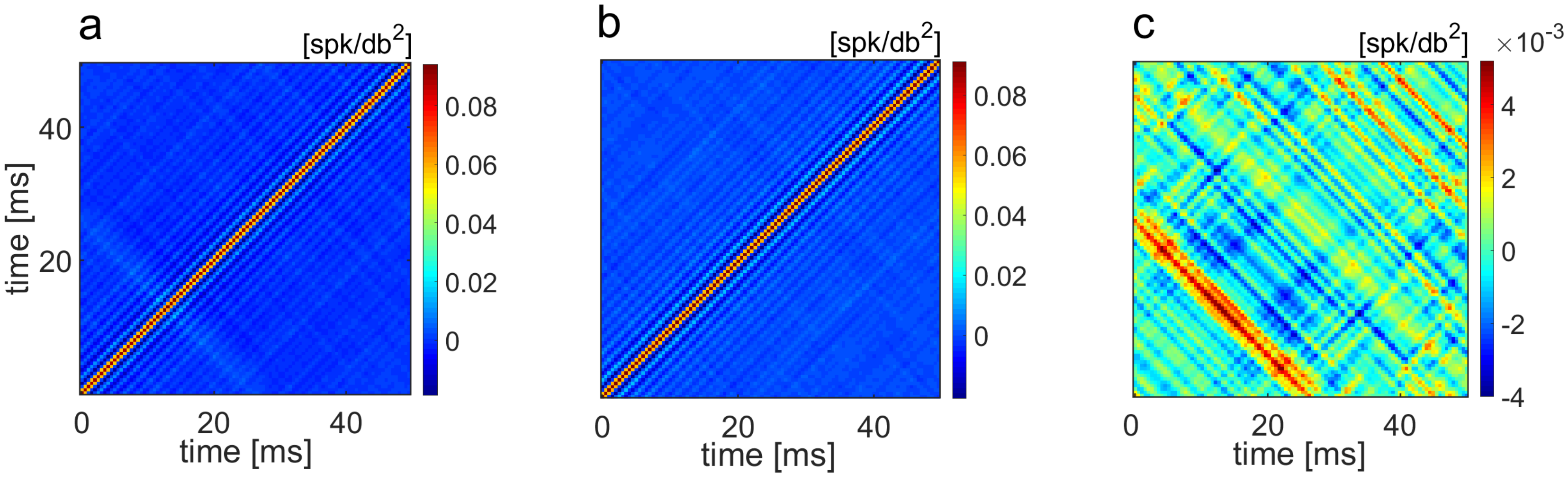}%
\caption[Generation of spike triggered covariance.]{The a) spike-triggered averaged outer product \(\mathrm{STC_{s}}\); the b) outer product obtained with jittered spike times \(\mathrm{C_{\mathrm{int}}}\); %
and the resultant c) % 
spike-triggered covariance STC; %
computed for one frequency channel of one neuron.}
\label{fig:subtractSTC}
\end{figure}

The STC is obtained from the second order Wiener kernel (Eq.~\ref{eq:WienerExpansion}). %
For each frequency channel \(f_{\mathrm{c}}\), the covariance matrix which is obtained from the whole stimulus \(C_{\mathrm{int}}(f_{\mathrm{c}})\), % 
is subtracted from the averaged outer product that is computed only from stimulus segments preceding spikes, %
\(\mathrm{STC_{s}(f_{\mathrm{c}})}\), %
\begin{eqnarray}\label{eq:stc}
		\mathrm{STC(f)}=\mathrm{STC_{\mathrm{s}}(f_{\mathrm{c}})}-\mathrm{C_{\mathrm{int}}(f_{\mathrm{c}})} % 
\end{eqnarray} 
% with the spike-triggered covariance
\(\mathrm{STC_{\mathrm{s}}}\) can be computed via: \\
\begin{eqnarray}\label{eq:STCspike}\mathrm{STC_{\mathrm{s}}(f_k)}=\frac{N}{T}\frac{1}{2(\frac{m_{\mathrm{db}}^2}{8})^2}\sum_{i=1}^{N} {m_{\mathrm{db}}}\frac{1}{N} \left(S(f_k,t_i-\tau)\!-\!\frac{{m_{\mathrm{db}}}}{2}\right) \otimes \left(S(f_k,t_i-\tau)'\!-\!\frac{{m_{\mathrm{db}}}}{2}\right)
\end{eqnarray}
with \(\mathrm{f_{\mathrm{c}}=f}\),~\(\mathrm{S_{\mathrm{DMR}}=S}\). %
The difference between a covariance and outer product is that for the covariance, for each (\(\tau_{1},\tau_{2}\))-combination the mean of each segments is subtracted before multiplication, whereas for the outer product it is not. The outer product can be used for the spike-triggered covariance analysis \citep{YamadaLewis1999}. %
The outer product is computed for each temporal DMR envelope segment of length \(\tau\) %
which elicited a spike at time \(\mathit{t_{i=1...N}}\), and averaged across all \(\mathit{N}\) obtained matrices, for each frequency channel, schematically shown in Fig.~\ref{fig:generateSTC}. % 
\\%
The number of spikes \(\mathit{N}\) varied from neuron to neuron, but was always higher or of the order of \(\mathrm{10^3}\), and the maximum time delay for each segment (\(\tau\,=\,\)50~ms) was the same as for the STA. The frequency resolution is \(\mathrm{\Delta f=}\)\!0.017~oct, with \(\mathrm{L=}\!\)220 frequency channels \(f_{\mathrm{c}}\). It is adjusted to sample densely (20\(\times\)) within the average ICC STRF bandwidth which is 1/3 octave. Each stimulus segment is multiplied by the mean intensity \(m_{\mathrm{db}}\!=\,\)30~dB, and centered. %
The \(\mathit{N}\) obtained outer products are averaged. % 
%The averaged \(\mathit{STC}\) ..
This averaged outer product is normalized by the mean firing rate \(\frac{N}{T}=N_{0}\), with the total recording time \(\mathit{T}\!=\,\)600~s, and with the squared standard deviation of the DMR sound \(\sigma ^2\,=\,\frac{{m_{\mathrm{db}}}^2}{8}\) \citep{Monty2002}. %
The so obtained matrices \(\mathrm{STC_{s}}\) from both trials A and B of the recording are averaged. %
Figure~\ref{fig:subtractSTC}a displays an example.% of an obtained 
\\
In order to obtain the correlation matrix which is only due to stimulus correlations that elicit a spike, the intrinsic stimulus correlations of the DMR envelope have to be removed. This is achieved by subtracting the average outer product of the whole DMR stimulus from \(\mathrm{STC_{\mathrm{s}}}\) (Eq.~\ref{eq:stc}). %
The DMR sound displays no global correlations, however, it displays short-term correlations. %
This has an effect on the computed spike-triggered outer product, see Fig.~\ref{fig:subtractSTC}a. Several paradiagonals of fading intensity with increasing distance from the diagonal are visible. % 
The effect of these short-term correlations becomes dominant for neurons which are highly feature selective (FS). For these neurons which fire almost exclusively in response to a specific combination of a temporal and spectral modulation, the local stimulus correlations captured in \(\mathrm{STC_{\mathrm{s}}}\) are predominant. They are not subtracted by the global mean, which is obtained by randomly selecting segments from the stimulus and averaging their outer products. %
To account for these local correlations, the spike times were added random temporal jitter from a range of %
-12,5~ms to 12,5~ms, because short-term correlations are still present within this range, however, envelope phase-locking is lifted. %
\newpage
The matrix describing the correlations of the stimulus \(\mathrm{C_{\mathrm{int}}}\) is expressed by a similar formula as the one for \(\mathrm{STC_{\mathrm{s}}}\), with the difference that the spike times %
have been added temporal jitter from a confined range%:\\
, \(\mathrm{ts_{\mathrm{i}}=t_{\mathrm{i}}+r_{\mathrm{i}}}\), %
\(\mathrm{r_{\mathrm{i}}\in[-12.5;12.5]}\)ms,%
\begin{eqnarray}\label{eq:STCshuffle}
%\(
\mathrm{C_{\mathrm{int}}(f)}=\frac{N}{T}\frac{1}{2(\frac{m_{\mathrm{db}}^2}{8})^2}\sum_{i=1}^{N} m_{\mathrm{db}} \left(S(f,ts_i-\tau)\!-\!\frac{m_{\mathrm{db}}}{2}\right)\otimes\left(S(f,ts_i-\tau)\!-\!\frac{m_{\mathrm{db}}}{2}\right)
%\)
\end{eqnarray}
The average outer product was computed from the DMR envelope for these jittered spike times and the obtained correlation matrix was subtracted from the spike-triggered correlation matrix \(\mathrm{STC_{\mathrm{s}}}\). The outer product for which temporal jitter has been added to the spike-times%
, \(\mathrm{C_{\mathrm{int}}}\), %
is shown in Fig.~\ref{fig:subtractSTC}b. % 
The paradiagonals are also visible %
here %
and the intensity range is the same as for the spike-triggered correlation matrix in Fig.~\ref{fig:subtractSTC}a. %
%\\
The only perceivable differences are the light orthodiagonals at around 20~ms in the spike-triggered correlation matrix. %
\\ 
Removing the intrinsic stimulus correlations from the spike-triggered outer matrix %
exposes %
the correlations which trigger the spiking of the neuron. The resultant correlation matrix is displayed in Fig.~\ref{fig:subtractSTC}c. %\\
This is the spike-triggered covariance, STC, analyzed in this paper. %
The paradiagonals due to short-term correlations are not perceivable anymore, but the previously light orthodiagonal is very pronounced. The intensity of this difference matrix is more than one order smaller than %
that of the two matrices in Fig.~\ref{fig:subtractSTC}a,b. %
\subsubsection*{Regularized STC}%
The significant %
noise-free \(\mathrm{STC(f_c)}\) for each frequency channel is obtained by summing the outer products of the significant singular vectors which are weighted by their singular values.
\begin{figure}[h!b]%[h!tb]
  \begin{minipage}[b]{0.49\linewidth}\includegraphics[width=0.87\linewidth]{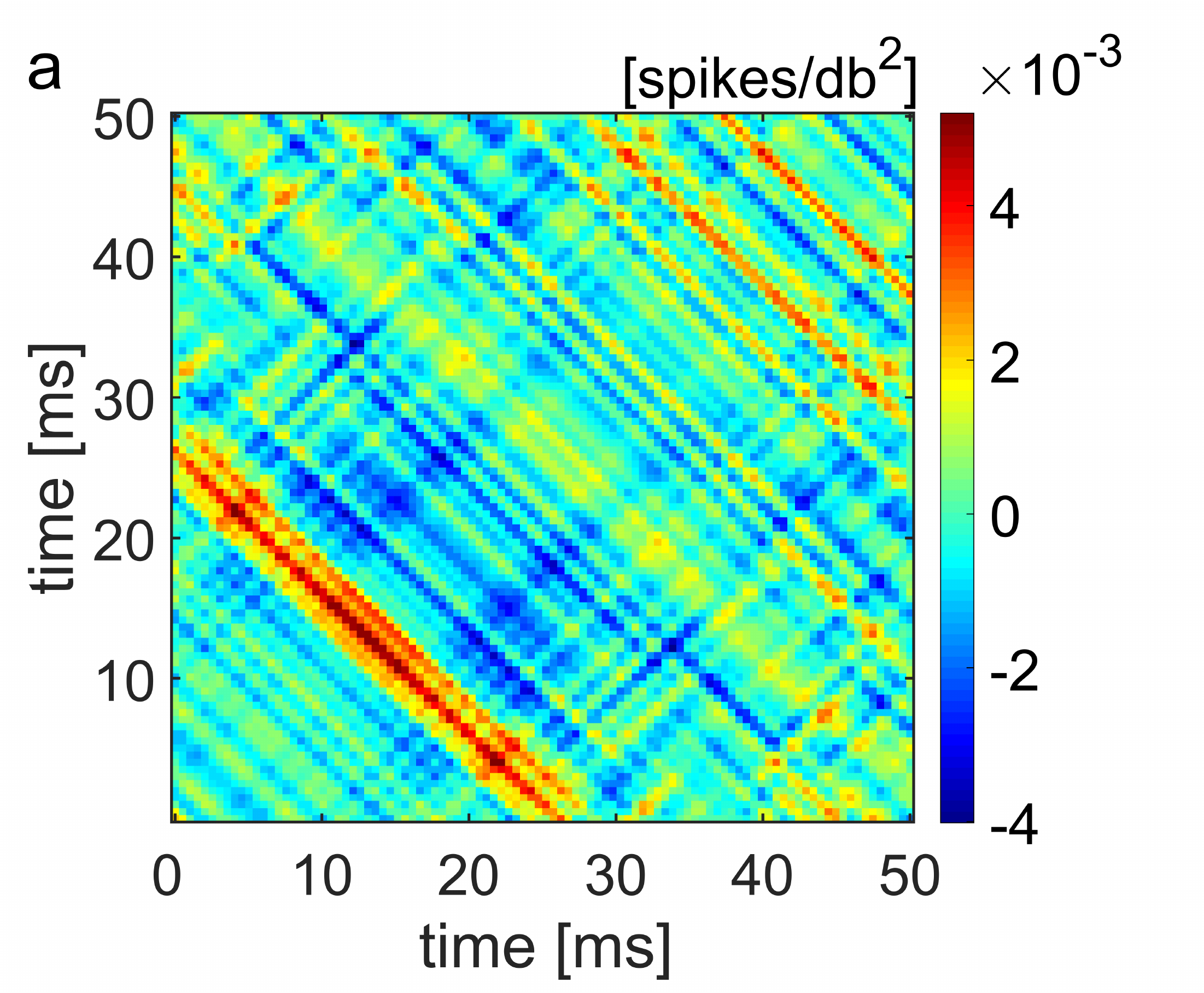}%98?
  \end{minipage}
    \hfill
  \begin{minipage}[b]{0.49\linewidth}\includegraphics[width=0.87\linewidth]{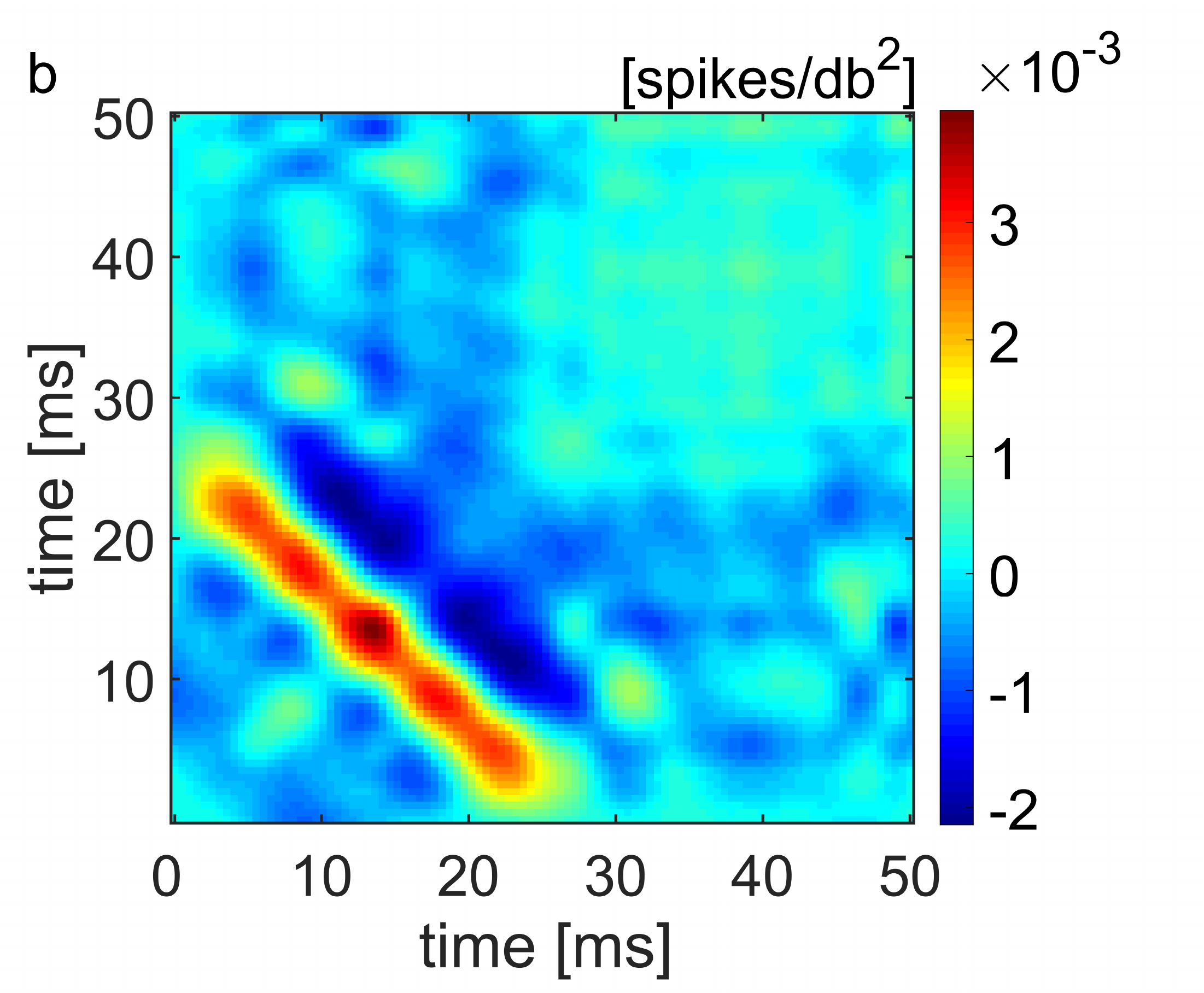}
\end{minipage}%
\caption[Reconstructed STC]{Reconstructed STC for a neuron and frequency channel (for which the singular value distribution is displayed in Fig.~\ref{fig:SSV1}); original STC (a); reconstructed STC, using all 5 significant vectors (b).}%
\label{fig:ReconSTC}
\end{figure}
\newpage
The regularized spike-triggered covariance \(\mathrm{STC_{\mathrm{R}}}\) %
for an \(\mathrm{STC}\) % 
with \(\mathit{n}\) significant singular values \(\mathit{\lambda_{j=1...n}}\) of sign \(\mathit{p}\) and singular vectors \(\mathit{\overrightarrow{S}_{j=1...n}}\) is given by \citep{Lewis2002}:\\
\begin{eqnarray}\label{eq:RecStc}
\mathrm{STC_{\mathrm{R}}} = \sum_{j=1}^{n}{\mathit{p}_{\mathrm{j}}\cdot\lambda_{\mathrm{j}}\cdot \overrightarrow{S}_{\mathrm{j}}\otimes \overrightarrow{S}'_{\mathrm{j}}}
\end{eqnarray}
The original and regularized STCs for one neuron and frequency channel are displayed in Fig.~\ref{fig:ReconSTC}. %
\subsubsection*{Spectrotemporal representation of STC}%
In order to enable comparing the energy in significant STCs to that in the linear STRF, the STA, a `STRF-like' representation is proposed. For each frequency channel, the time-dependent highest significant singular value are plotted. If no significant values were found for this frequency channel, the vector equals a zero-vector. This yields a matrix of energy values in dependance of the frequency channel and the time delay; the representation is of the same dimensions as the linear STRF.
The neuron's filtering properties are also indicated by the STRF; neurons that display interleaved patterns of excitatory % 
and inhibitory %
subfields in their STRFs generally have bandpass response characteristics. If such interleaved patterns are absent, the neurons generally have lowpass response characteristics. Neurons which are not phase-locking to the envelope, no statistically significant STRF can be derived because responses to different spectrotemporal features average out. 
\subsubsection*{Significance testing of spike-triggered covariance}%
\begin{figure}[h!bt]\centering
  	    \includegraphics[width=0.65\linewidth]{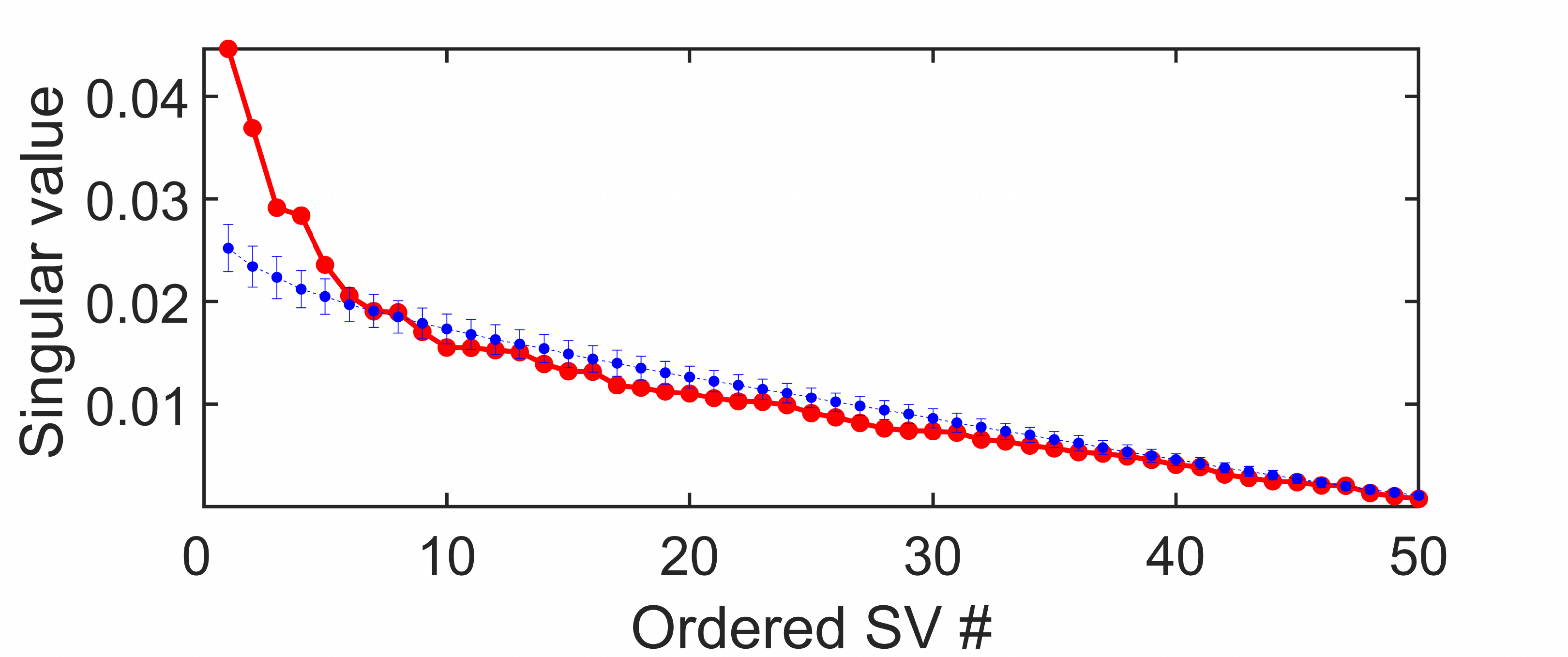}%
\caption[Significance testing.]{Significance testing. Ordered distribution of singular values which are derived from the STC (Fig.~\ref{fig:subtractSTC}a) for the frequency carrier \(\mathrm{f = 2.6~oct}\). Values (\mycirc[red]\small{) exceeding the estimated noise level (}\mycirc[blue]\small{) by at least \(\mathrm{\Theta}=1.6\;\sigma\) (\(\mathit{p}\)\(=\)0.05) standard deviations are considered significant.}}
\label{fig:SSV1}\end{figure}

Significance was assessed using Student's \(\mathit{t}\)-test for normal distributions, and the Wilcoxon-Mann-Whitney test for comparison of non-normal distributions (significance level \(\mathit{\alpha=}\)0.05); the \(\mathit{p}\)-value is given.\\
In order to determine if the computed STC matrix of stimulus correlations was indicative of neuronal preferences, to test if it was statistically significant above noise, significance testing was performed. Singular value decomposition of the STC matrices for each frequency channel is carried out, and yields distributions of singular values (SV). %
To estimate the noise level from the recorded data the following is performed, bootstrapping: an STC is computed using Eq.~\ref{eq:stc} from randomly chosen 1~min-long segments of the 10~min spike-train recording. From this matrix \(\mathrm{STC_{\mathrm{boot}}}\), the singular values and vectors are derived, and averaged for both trial sets A and B. This procedure is repeated 100~times for different segments, and the mean and standard deviation of these 10 singular value distributions were taken. This yields an estimate of the noise level, the estimated error with a standard deviation. The singular values of STC and % 
\(\mathrm{STC_{\mathrm{boot}}}\) (noise) are ordered according to decreasing value. The distributions of singular values from the STC and the estimated noise, for one frequency channel (\(\mathrm{f = 2.6~oct}\)) of a neuron are given in Fig.~\ref{fig:SSV1}. %
The spike-triggered covariance is considered significant if its singular values exceed the noise by at least \(\mathrm{\Theta}=1.6\;\sigma\) standard deviations (\(\mathit{p}\)\(=\)0.05). % 
\newpage
The significant vectors and their corresponding values were used to reconstruct a STC for which the noise is removed, \(\mathrm{STC_{\mathrm{R}}}\), Eq.~\ref{eq:RecStc}. %
We estimated significant STCs if they displayed at least one significant singular vector. The analysis has additionally been performed using a higer threshold, \(\mathrm{\Theta}=3\;\sigma\) (\(\mathit{p}\)\(=\)0.001). Unless stated otherwise, only significant STCs are further analyzed in this work.%
\section*{Results}%
The described analysis was applied to n\(=\)178 single neurons. %   
Significant singular values (SV) and thus significant STCs for at least one frequency channel were found in n\(=\)132 neurons (75\(\mathit{\%}\)). These are further described in the following. For an elevated threshold of \(\mathrm{\Theta}=3\;\sigma\) when testing for significance, n\(=\)102 (57\(\mathit{\%}\)) displayed significant \(\mathrm{STC}\).
\subsubsection*{Significant values across frequencies}%
\begin{figure}[h!btp] 
 \centering	
  \begin{minipage}[b]{0.8\textwidth} 
   \includegraphics[width=0.9\linewidth]{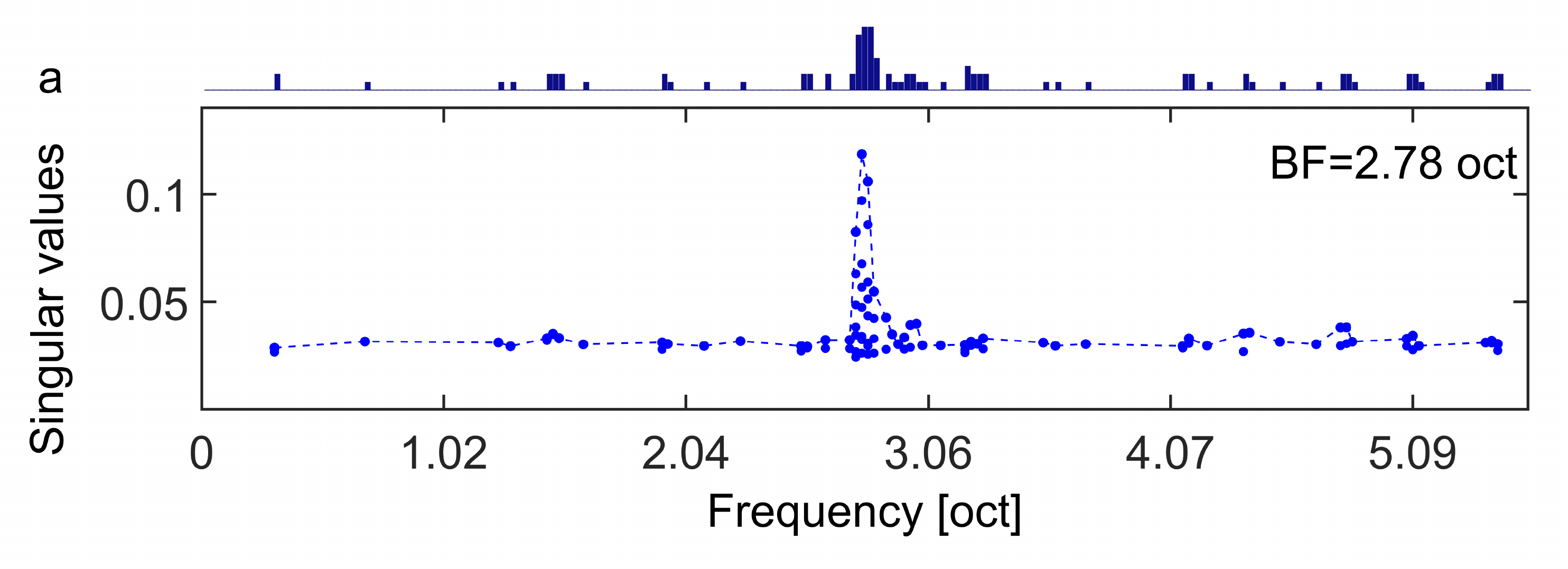}%.eps
  \end{minipage} 
  \vfill
  \begin{minipage}[b]{0.8\textwidth} 
   \includegraphics[width=0.9\linewidth]{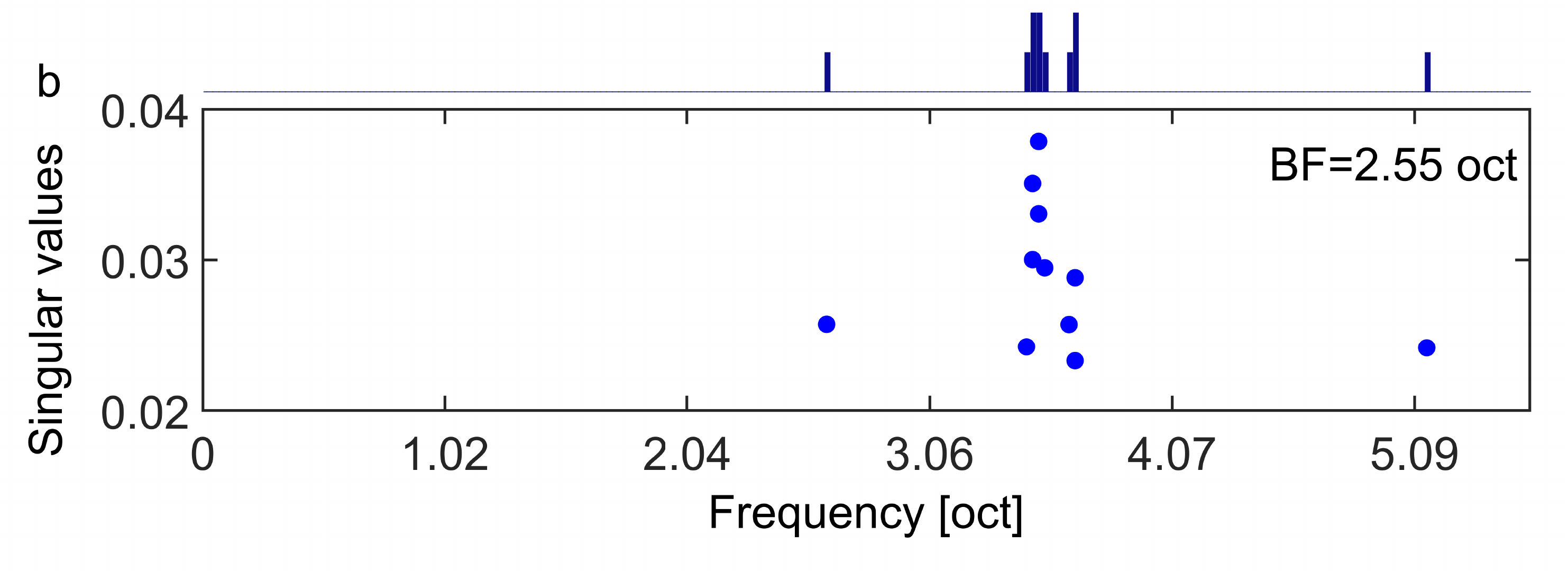}%.eps
  \end{minipage} 
  \vfill
  \begin{minipage}[b]{0.8\textwidth} 
   \includegraphics[width=0.9\linewidth]{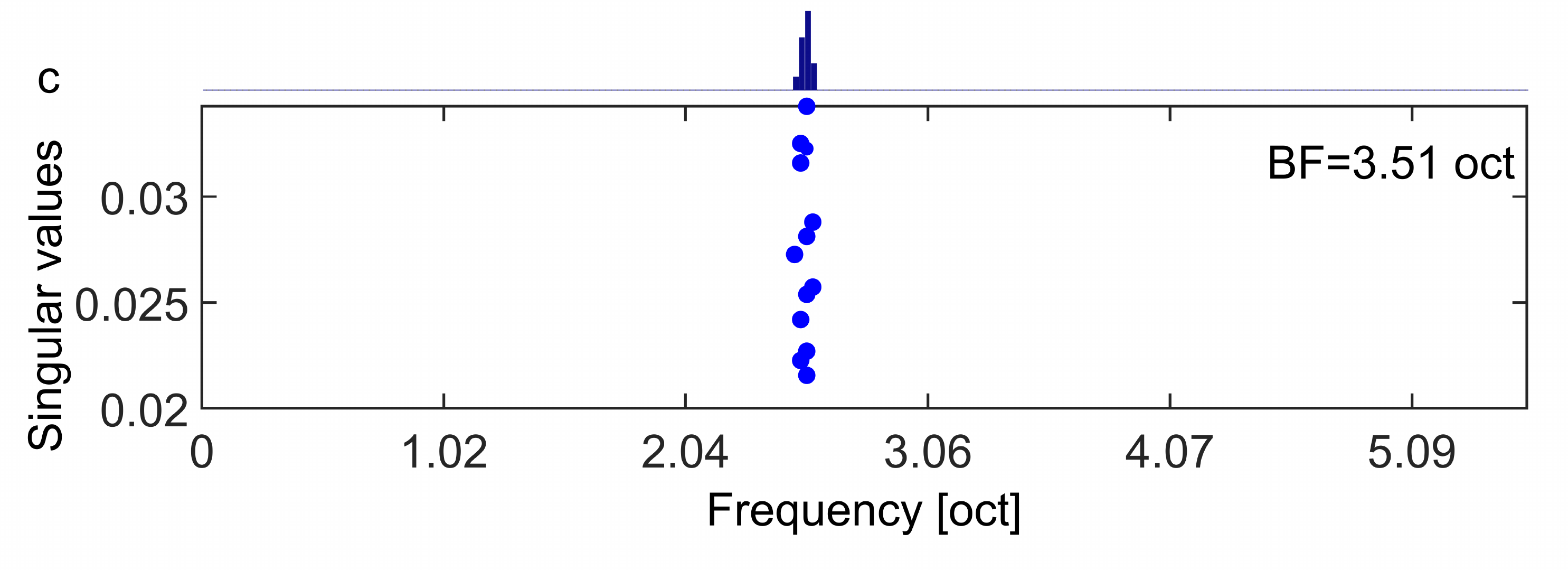}%.eps
  \end{minipage}   
\caption[Distribution of significant values across frequency channels.]{Distribution of significant singular values across frequency channels. The distributions of three different neurons a), b) and c) are given (\(\mathrm{\Theta}=1.6\;\sigma\),~\(\mathit{p}\)\(=\)0.05). The blue dashed line is a guide to the eye. The bar histogram on top of each graph shows the relative number of significant values for each frequency channel.}
\label{fig:SVdistr}
\end{figure}

\begin{figure}[bhp]\centering	
   \includegraphics[width=0.65\linewidth]{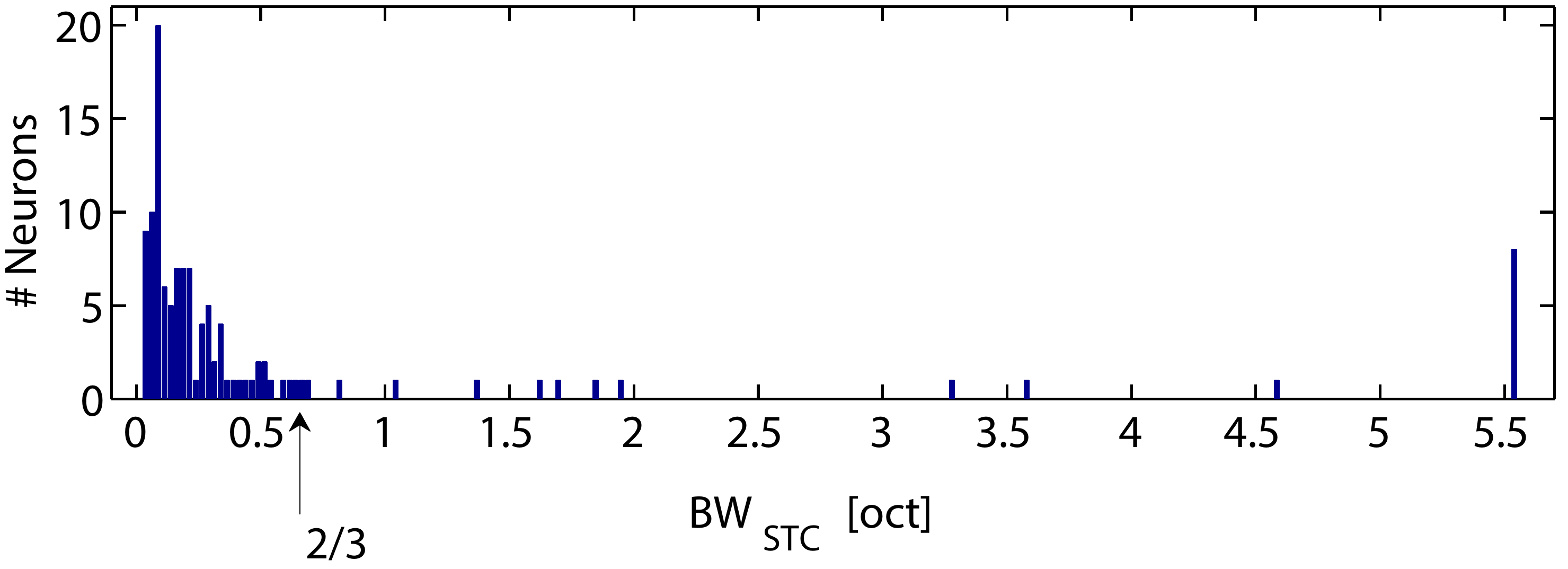}
\caption[Distribution of STC bandwidth.]{Distribution of significant STC bandwidth \(\mathrm{BW}_{\mathrm{STC}}\). Coherent spectral bandwidth for which for which the STCs are significant, for all neurons. The width of this coherent frequency range around the best frequency% 
lies within 2/3 octave for most of the neurons (\(\mathrm{\Theta}=1.6\; \sigma\),~\(\mathit{p}\)\(=\)0.05).}
\label{fig:BWstc}
\end{figure}

The values and the total number of singular values (SV) derived from the STCs varied across frequency channels. In Fig.~\ref{fig:SVdistr}, the distribution of significant singular values across all 220 frequency channels is shown for three neurons as an example. In most cases, the highest significant singular values and also the largest quantity of singular values were obtained from STCs at and around the best frequency. STCs of other frequency channels still yielded significant values, but of smaller magnitude and quantity (Fig.~\ref{fig:SVdistr}a). For some neurons, only the STC of one frequency channel, which was at or close to the best frequency, yielded significant values (Fig.~\ref{fig:SVdistr}b). In a few cases, only a few significant values were found, at the best frequency, but also at frequency channels more than one octave away (Fig.~\ref{fig:SVdistr}c).%
\newpage
%
%\begin{figure}[h!tbp] \centering	
   %\includegraphics[width=0.65\linewidth]{BFmaxDiffb.pdf}%BFmaxDiff.eps
%\caption[STCs at the best frequency have the highest singular values.]{For each neuron, the absolute spectral difference (in octaves) between the best frequency \(\mathrm{BF}_{\mathrm{STRF}}\), and the frequency channel of the STC displaying the highest singular value, \(\mathrm{BF}_{\mathrm{STC}}\), are shown. The STCs at the best frequency channel have the highest singular values.% 
%}
%\label{fig:BFmaxDiff}
%\end{figure}%
%%
In most cases, the highest singular values and often the largest quantity of SVs was found for the STC at the best frequency (BF) channel. % 
%
%,~Fig.~\ref{fig:BFmaxDiff}. %
%
For 92 neurons out of 132, %
the best frequency yields the highest singular value, and for
97$\%$ of the neurons, the frequency channel with the highest singular value and the best frequency differ by less than half an octave (\(\mathrm{\Theta}=1.6\;\sigma\),~\(\mathit{p}\)\(=\) 0.05). %
The dominant STCs are at the best frequency, and the magnitude of the singular values generally falls off symmetrically to both frequency sides of the BF. The width of the coherent, gap-less frequency range around the best frequency which displays significant singular values was computed for each neuron. This absolute spectral width is small for example for the neuron described in Fig.~\ref{fig:SVdistr}c. For 84\(\%\) of the neurons, this coherent bandwidth around the BF equals or is less than 2/3 of an octave, see~Fig.~\ref{fig:BWstc}. %
The singular values differ across frequency channels, as do the STCs from which they were derived. These differ not merely in magnitude but in their shape.%
%
%\newpage
\subsubsection*{Significant STCs}%
\begin{figure}[h!tb]\centering
    \includegraphics[width=0.99\linewidth]{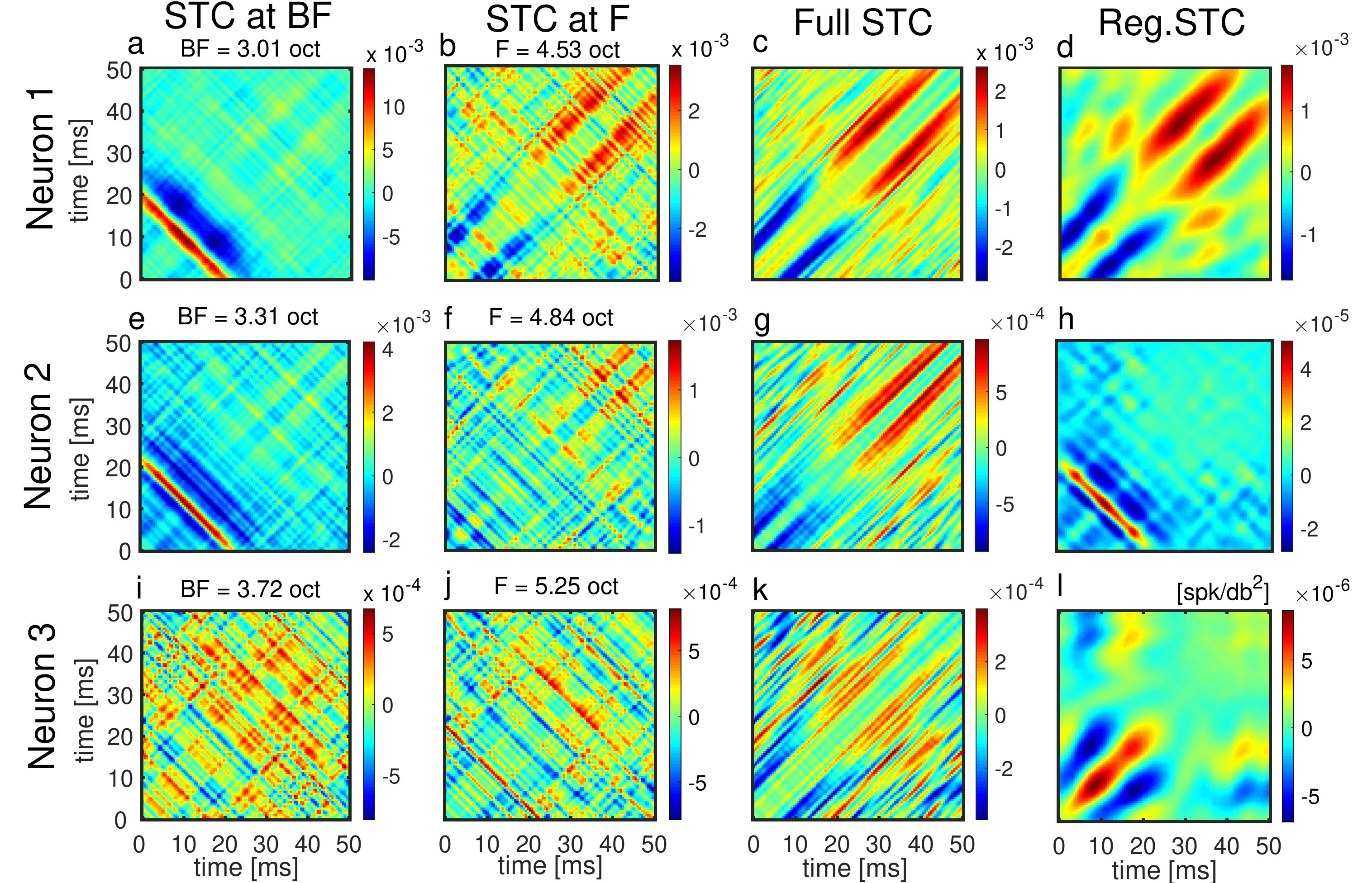}%exp.eps
\caption[Examples of spike-triggered covariances.]{Examples of spike-triggered covariances %
for three neurons. Neuron 1~(a-d), Neuron 2~(e-h), Neuron 3~(i-l). For each neuron, the spike-triggered covariance at the best frequency (a,~e,~i) and at a frequency channel more than one octave away (b,~f,~j) are displayed. The averaged STCs across all frequency channels are given, the `Full STC' (c,~g,~k) and its regularized STC, which is constructed only from significant STCs, the `Regularized STC' (d,~h,~l). For all \(\mathrm{\Theta}=3\;\sigma\),~(\(\mathit{p}\)\(=\)0.05), except for h), \(\mathrm{\Theta}=1.6\;\sigma\) (\(\mathit{p}\)\(=\)0.001).}%
\label{fig:stc1}
\end{figure}

\begin{figure}[th!]\centering%[h!t]
    \includegraphics[width=0.95\linewidth]{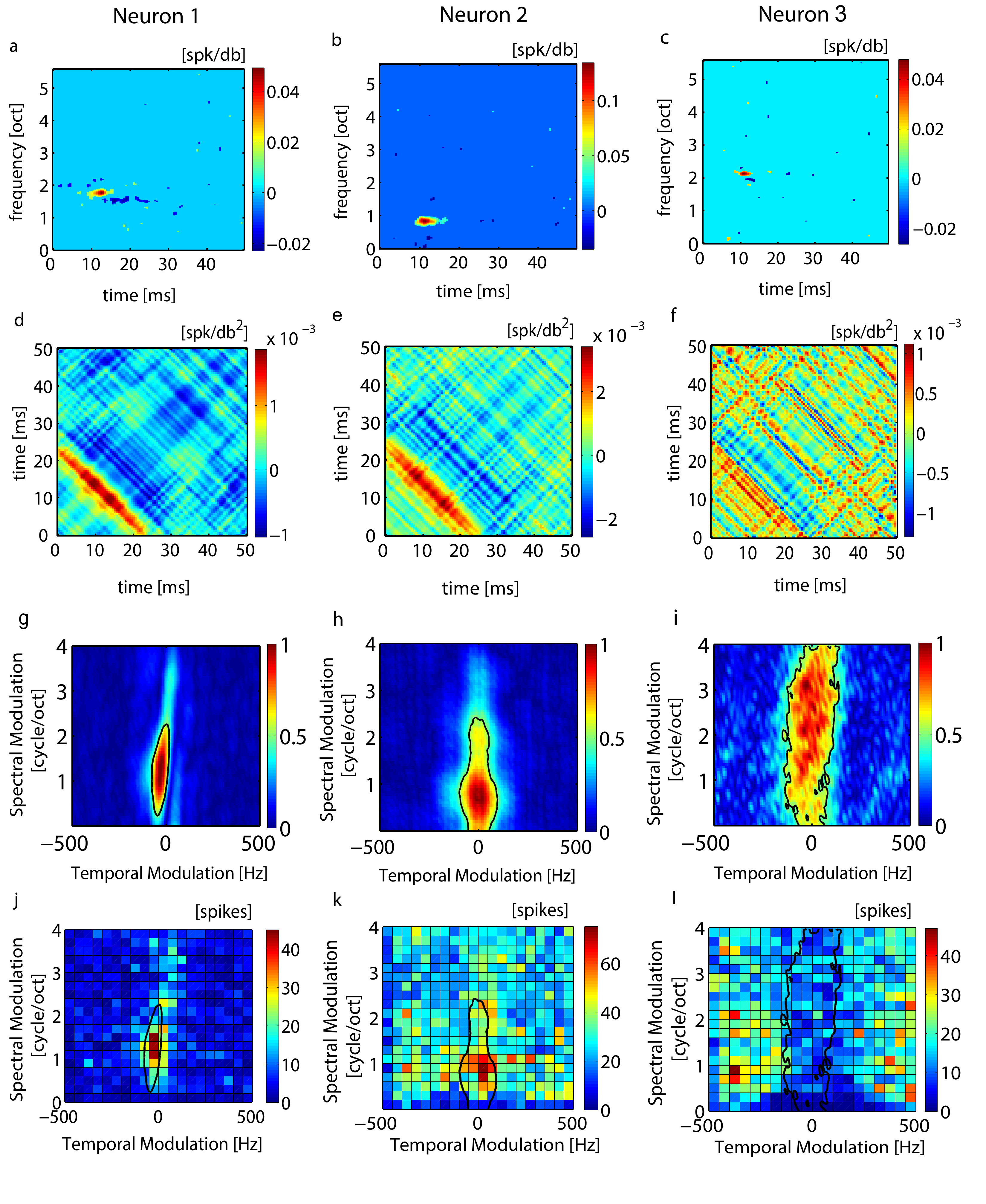}
\caption[Comparison of STC to other descriptors of neuronal preferences.]{Comparison of STC to other descriptors of neuronal preferences. For Neuron 1,~2 and 3, the significant linear spectrotemporal receptive field STRF (a-c), the spike-triggered covariance STC at BF (d-f), the ripple transfer function RTF (g-i) and the conditioned response histogram CRH (j-l) are displayed (\(\mathrm{\Theta}=3\;\sigma\),~\(\mathit{p}\)\(=\)0.001).}
\label{fig:comp}
\end{figure}

STCs vary across frequency channels and across neurons (Fig.~\ref{fig:stc1}). Figures~\ref{fig:stc1}a,e show representative STCs with orthodiagonals at the best frequency. Stimulus correlations with maximum delays of 30~ms elicit enhanced spiking activity. The stimulus correlations follow an excitatory-inhibitory pattern (Fig.~\ref{fig:stc1}a) or an inhibitory-excitatory-inhibitory (Fig.~\ref{fig:stc1}e) pattern. %
However, STCs which display neuronal preferences for stimulus correlations with a constant time delay at around 10~ms are representative %
mostly for STCs off the best frequency (Fig.~\ref{fig:stc1}b). The displayed STC shows a temporal series % 
of inhibitory and excitatory stimulus correlations with a constant delay. When averaging the STCs across all frequency channels, these stimulus correlations are dominant (Fig.~\ref{fig:stc1}c,g), and are in some cases also pronounced when only considering significant STCs (Fig.~\ref{fig:stc1}d) but also averaged out in other cases (Fig.~\ref{fig:stc1}h). %
The reconstructed STC from all frequency channels of Neuron 3 (Fig.~\ref{fig:stc1}l) displays preferences for stimulus correlations with constant delay and suppressive spiking activity for constant delay correlations of 10~ms.%
\newpage
In some cases, STCs do not show a pronounced pattern, but yield significant singular values (Fig.~\ref{fig:stc1}f,i,j).
In order to investigate whether %
significant STCs can be predicted by other descriptions of the neurons, these were compared to the emergence of significant STCs. %
For three neurons, the spike-triggered covariances at the best frequency are displayed together with their significant STRFs, the derived ripple transfer functions (RTF) and conditioned response histograms (CRH) in Fig.~\ref{fig:comp}.%
\\
The CRH counts the occurrences of spikes to a particular parameter combination. The RTF also shows preferences for parameter combinations, however, it is derived from the linear STRF, and might not show the full preference spectrum of the neuron, since the STRF could be averaging out preferences. A large deviation between the CRH and RTF could indicate nonlinear response properties. Significant STCs occurred for neurons that showed a large overlap of RTF and CRH (Neuron 1,~Fig.~\ref{fig:comp}d,g,j), but also for neurons for which the overlap was not very good (Neuron~2,~Fig.~\ref{fig:comp}e,h,k). %
In the example of Neuron 2, the CRH is %
diffuse and shows no preference, the RTF, however, does show bandpass tuning in the temporal domain and lowpass tuning in the spectral domain. These were the two cases, for which mostly significant STCs were encountered: good overlap of RTF and CRH, or no good overlap due to a diffuse CRH. %\\
For neurons which yielded no significant STC, the RTF and CRH usually showed no overlap, either they showed a complementary behavior with no overlap in the parameter space (Neuron 3,~Fig.~\ref{fig:comp}f,i,l), or the CRH was diffuse.%
\subsubsection*{Comparison of significant linear STRF and STC}%
\begin{figure}[h!tbp]\centering
  \begin{minipage}[b]{0.80\linewidth}\includegraphics[width=0.87\linewidth]{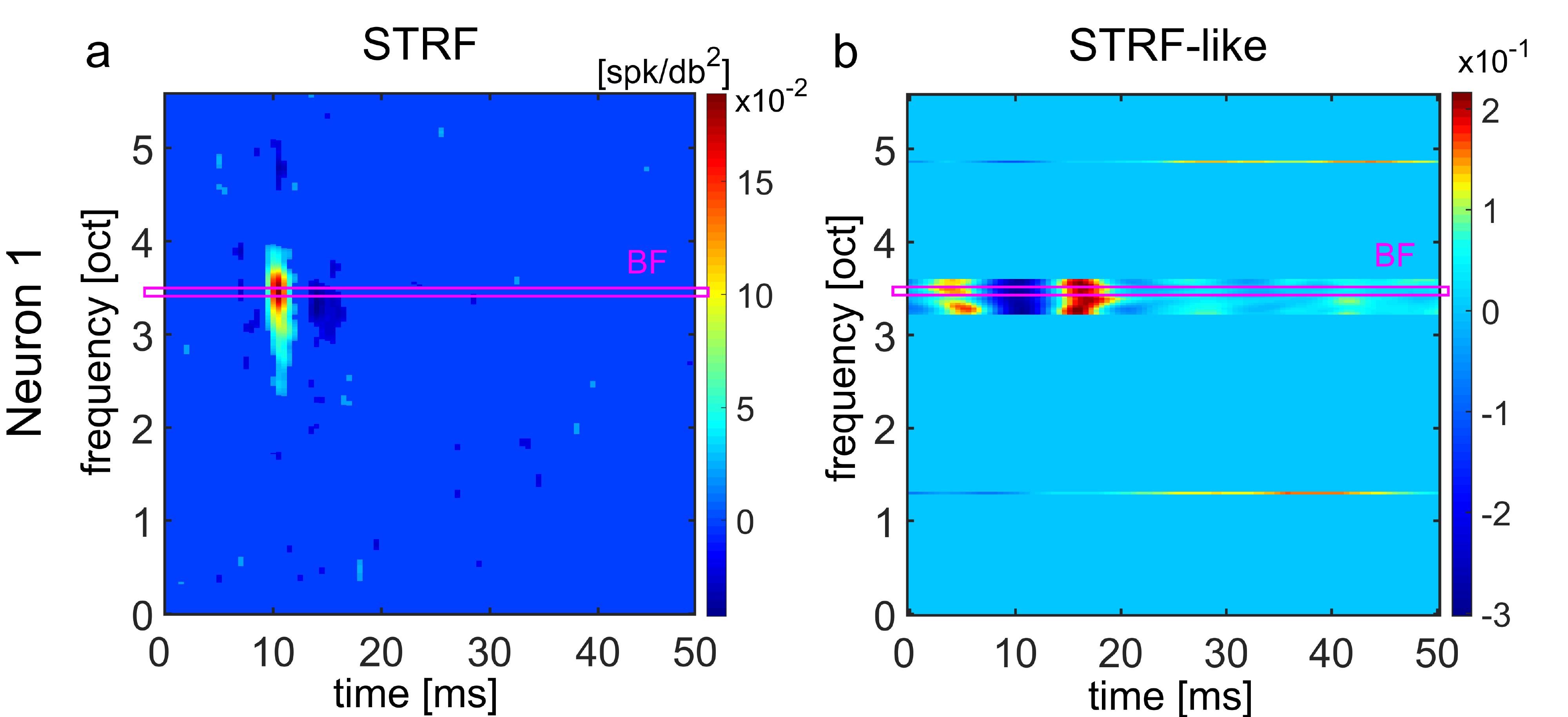}%
  \end{minipage}
    \vfill
  \begin{minipage}[b]{0.80\linewidth}\includegraphics[width=0.87\linewidth]{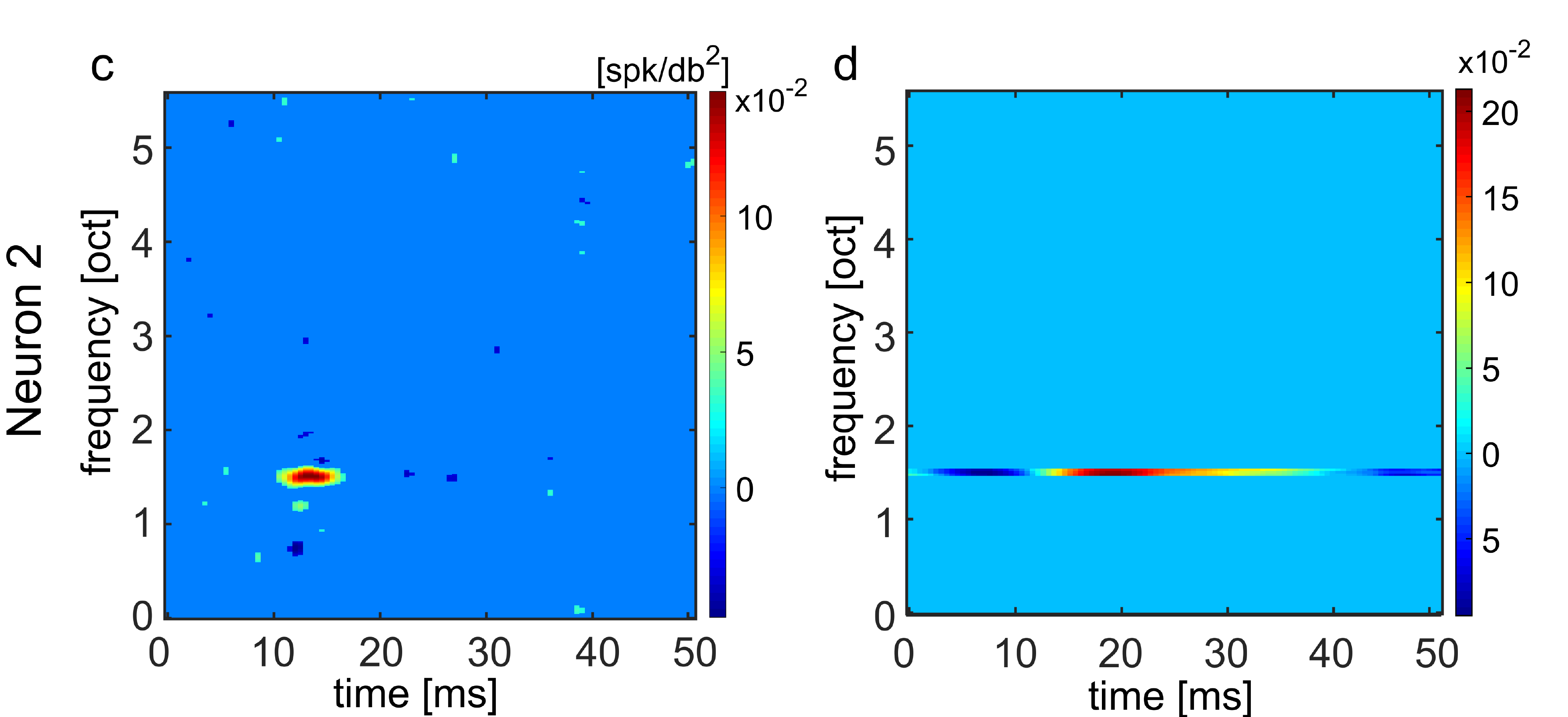}
\end{minipage}%
\caption[Comparison of linear STRF and STC.]{Comparison of energy in significant linear STRFs and STCs. The significant linear STRF is compared to the first significant vectors of each frequency channel using a `STRF-like' representation, for Neuron 1 (a,b), with highlighted best frequency channel BF, and Neuron 2 (c,d),~\(\mathrm{\Theta}=3\;\sigma\),~\(\mathit{p}\)\(=\)0.001).}
\label{fig:strflike}\end{figure}

We showed that dominant frequency regions are very similar for the linear STRF and the STC. To test whether this is also true for energy in the temporal domain, the linear STRFs are compared to the temporal representation of the STC in each frequency channel. For the comparison we use a spectrotemporal representation of the STC, the `STRF-like' representation. Examples for two neurons are given in Fig.~\ref{fig:strflike}.% 
\newpage
For Neuron 1 the STRF displays a broad bandwidth and an excitatory-inhibitory temporal energy pattern with a delay time of about~10~ms (Fig.~\ref{fig:strflike}a). The bandwidth of the STC is smaller and contained within the bandwidth 
of the %
STRF. In the spectrotemporal representation of the STC, several significant vectors around the BF display energy with an excitatory-inhibitory-excitatory temporal pattern centered also at a delay of 10~ms (Fig.~\ref{fig:strflike}b). Thus, the tuning is opposite in these two receptive fields, and the nonlinear contribution captured by the STC could be modulatory %
as proposed previously for auditory neurons in the midbrain of songbirds \citep{WoolleyExtraTuning2011}. %
In the second example only one significant vector is present, which is located at the best frequency. Neuron 2 shows an excitatory region at about 12.5~ms delay (Fig.~\ref{fig:strflike}c). The singular vector of the STC displays an inhibitory-excitatory temporal pattern with an excitatory region at a delay time of 20~ms (Fig.~\ref{fig:strflike}d). Thus, also in this case, the temporal modulations do not coincide, but are shifted. The linear STRF and STRF-like representation of the STC show enhanced (excitatory and suppressive) energy in similar spectral and temporal regions. %
\newpage
\begin{figure}[h!btp]\centering
    \includegraphics[width=0.85\linewidth]{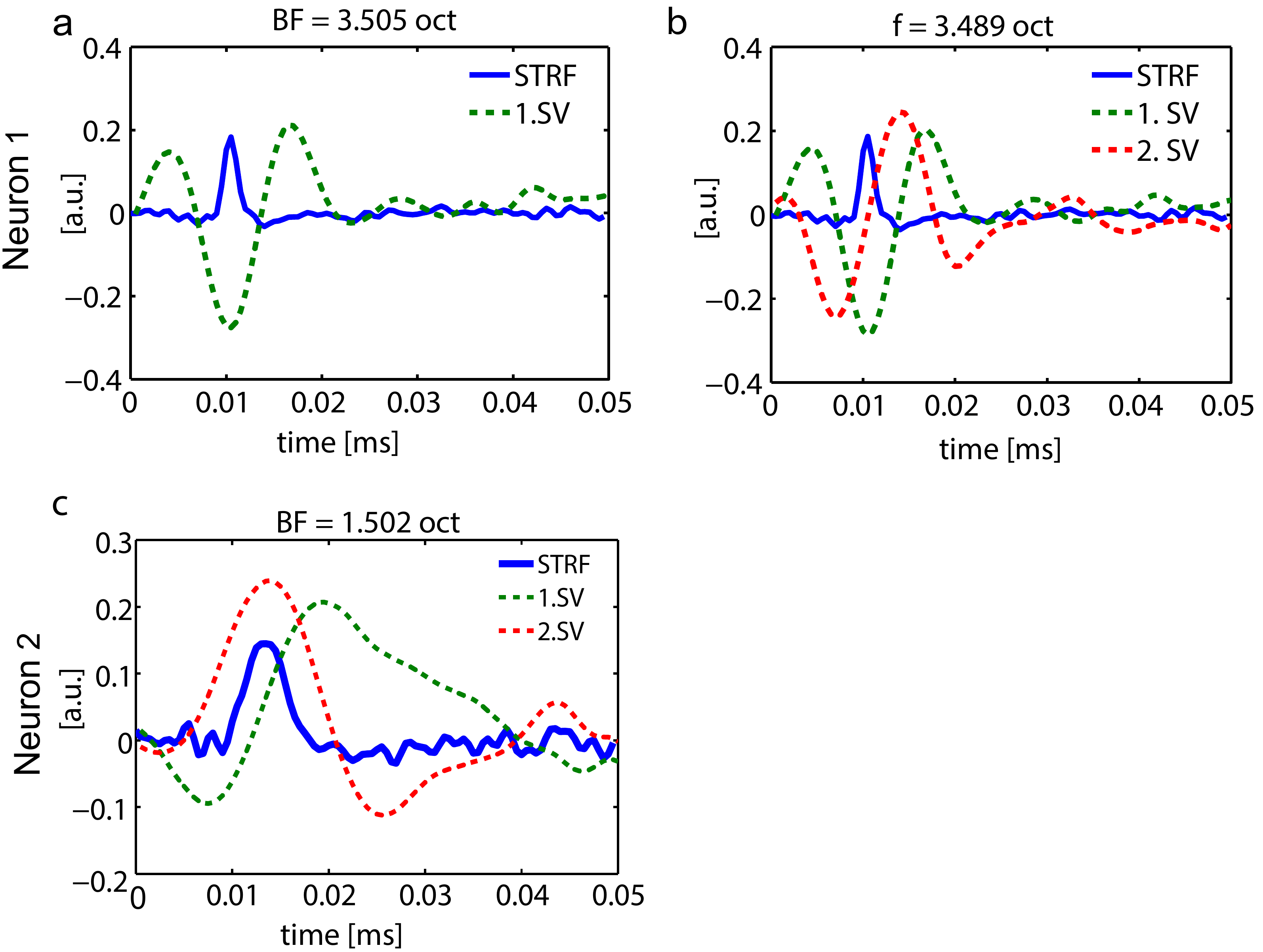}
\caption[Temporal course of significant vectors.]{Comparison of temporal course of significant STRF and of first (and second) significant singular vectors (SV) for the two neurons displayed in Fig~\ref{fig:strflike}. a) Neuron 1: temporal courses of linear STRF and first significant vector, at the best frequency. %
This corresponds to the highlighted area in Fig.~\ref{fig:strflike}a,b
; b) Neuron 1: temporal courses of linear STRF and first and second significant vector, at a frequency channel near the best frequency; c) Neuron 2: temporal courses of significant STRF and first and second significant vector, at best frequency. Neuron 2 only displays significant values at the best frequency.}
\label{fig:kernel}\end{figure}%
\newpage
Significant singular vectors and the temporal courses of the corresponding frequency channels in the linear STRF are shown in Fig.~\ref{fig:kernel}. The opposing tuning of the temporal course of the STRF at the best frequency and of the first significant vector is shown in Fig.~\ref{fig:kernel}a. For a neighboring frequency the tuning is very similar; the second significant vector is shifted by a few milliseconds relative to the first one (Fig.~\ref{fig:kernel}b). In the example of Neuron 2, % 
the second significant vector coincides temporally with the excitatory region of the linear STRF.%
\begin{figure}[h!]\centering
 \includegraphics[width=0.78\linewidth]{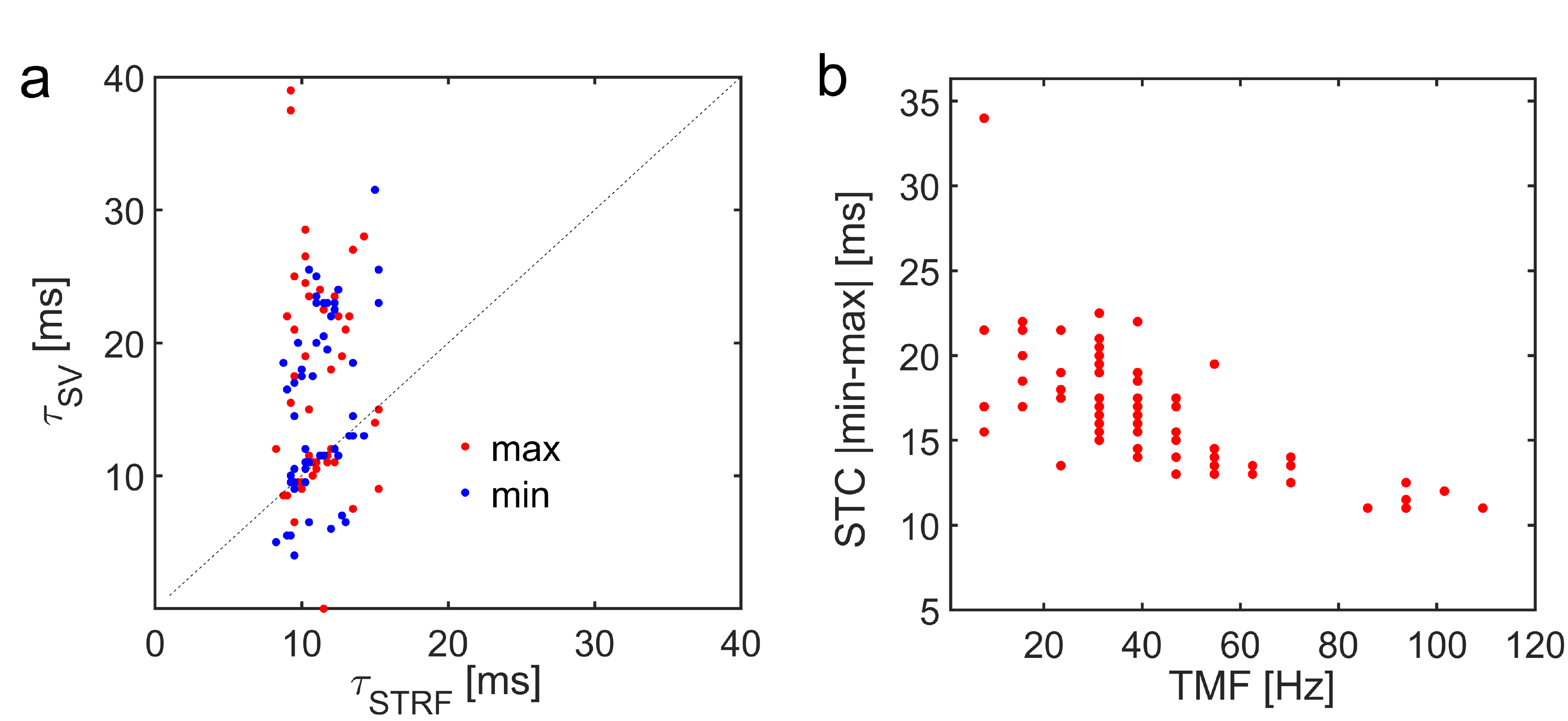}
   \caption[Temporal tuning of STRF and STC.]{Temporal tuning of linear STRF and first singular vector of STC at the best frequency, across all neurons. a) scatter plot of delay times from STRF \(\mathrm{\tau}_{\mathrm{STRF}}\), and from STC,\(\mathrm{\tau}_{\mathrm{STC}}\), with first extremum either a minimum %
(\mycirc[blue]\small{) or a maximum (}\mycirc[red]\small{); b) width in temporal tuning of STC vs. neuron's best temporal modulation frequency (TMF).}}
  \label{fig:delay}
\end{figure}%
\\
The relationship between the delay time, as determined from the linear STRF and the delay of the first extremum (either minimum or maximum value) of the first significant singular vector at the best frequency %
is displayed for all neurons in Fig.~\ref{fig:delay}a. %
No significant differences exist for whether the first extrema were minima or maxima. %
For 47\(\%\) of the neurons %
these two delays matched (difference \(<\)~1.1~ms), for both, mimima and maxima, hence the tuning of the singular vector was the same or opposite to the STRF. For almost all other neurons, the delay of the STC was higher than that for the respective STRF, with differences ranging mostly between 5 and 30~ms. This was true for STCs with same and opposite tuning. For a small percentage of the neurons (10~\(\%\)), delay times of the STC were smaller than the ones found with the linear STRF by about 5~ms. % 
\begin{figure}[bh!tp]\centering
 \includegraphics[width=0.78\linewidth]{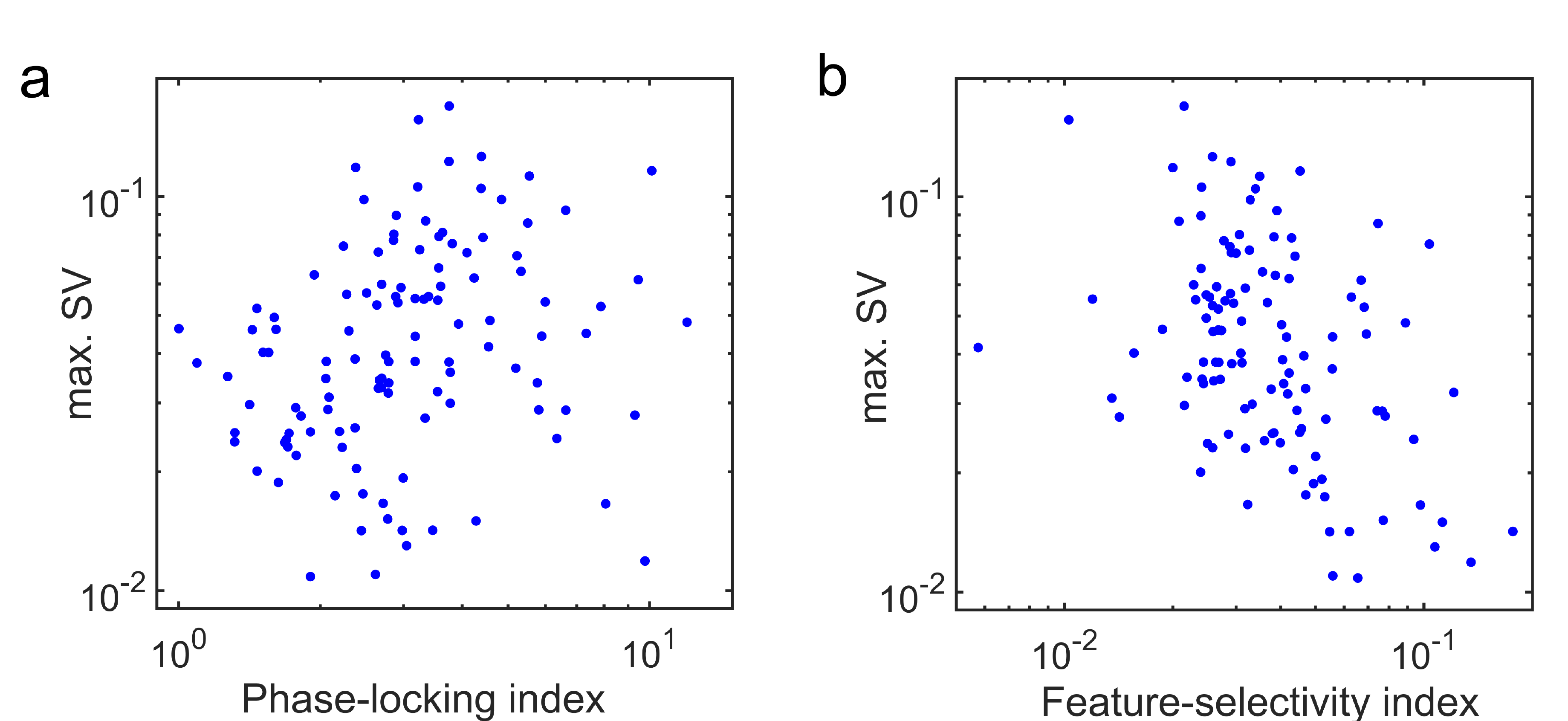}%
    \caption[Relationship between selectivity and degree of nonlinearity.]{
		Measure of nonlinearity in relation to the phase-locking (PLI) and feature-selectivity index (FSI), for all neurons. %
The scatter plots show the highest singular value at the best frequency against a) the PLI; b) FSI.}
  \label{fig:prop}
\end{figure}%
\\
To further investigate temporal tuning, the temporal width between extremes in the temporal course of the singular vectors was compared to the best temporal modulation frequency of the neuron (TMF) %
, Fig.~\ref{fig:delay}b. %
There is a tendency of larger temporal widths corresponding to low TMFs, and vice versa. %
Previous studies have suggested that nonlinear receptive fields might contribute to an increase of the neuron's feature selectivity \citep{GallantRF2002}. The highest significant value at the best frequency was taken as a measure in order to quantify the nonlinear processing of a neuron.%
\newpage This measure was compared against the degree of phase-locking, using the neuron's phase-locking index (PLI), %
see Fig.~\ref{fig:prop}a, %
and against the feature-selectivity index (FSI)%
, Fig.~\ref{fig:prop}b. % 
Between the phase-locking index %
 and the degree of nonlinear responses no clear relationship is visible. % found?
A tendency of higher singular values corresponding to low feature selectivity, and vice versa is visible.
\section*{Discussion}%
We demonstrated that more than half of the neurons (even 75\( \%\) for \(\sigma=\)1.6) in the mammalian ICC show significant preferences to stimulus correlations. %
Preferences to envelope correlations in the ICC were probed using the spike-triggered covariance in response to the dynamic moving ripple sound stimulus. The spike-triggered covariance method was used for detecting nonlinear response properties, because it allows capturing all relevant stimulus features.\\
The nonlinear preferences vary across neurons and across frequency channels. The STC shapes are as diverse and rich as the linear STRFs. A nonlinearity present in more than half of the neurons was found for stimulus correlations with maximum time delays of about 30~ms. Dominant nonlinear response properties are almost exclusively present at the best frequency of the neuron, and comparing the significant linear STRF and the first significant vectors of the covariance reveals a large spectral overlap, but also reveals shifted temporal courses. About half of the neurons displayed the same tuning pattern and the other half displayed opposite tuning pattern. Same tuning could enhance the response and the opposite contribute to further tuning. The nonlinear contribution from the STC could be modulatory as proposed previously for auditory neurons in the midbrain of songbirds \citep{WoolleyExtraTuning2011}.\\ %
%
%
%Controversy-nonlinearities.
It has been suggested previously that the processing of ICC neurons are well described by a linear receptive field \citep{SharpeeRFdim2012}, however, another study suggested that ICC neurons are tuned to multiple (also nonlinear) features \citep{AndoniPollakFMselecSTC2011}. %
Our findings are supported by the work of Andoni and Pollak \citep{AndoniPollakFMselecSTC2011}. 
In an information-theoretic approach, they showed that the most informative features were selective for the spectral motion of frequency modulation sweeps and suggest %
that two mechanisms exist which induce this selectivity, and which apply respectively to half of the neurons. These mechanisms were proposed earlier for neurons in the visual system \citep{Adelson1985}. The first mechanism for the processing of neurons consists of two linear filters with a quadratic phase shift whose output is squared and summed. In the second proposed mechanism neurons are tuned for opposing directions, which either increase or suppress the spiking activity. This mechanism consists of two linear filters with opposite orientations and a spiking response which corresponds to the difference between their squared output \citep{Adelson1985}. % 
This is in agreement with our findings and further supports that neurons are tuned to multiple spectrotemporal features and that selectivity for natural communication calls can already be observed at the level of the inferior colliculus.\\
The found % 
nonlinearities could be already present in the input to the ICC or they could originate from nonlinear processing of ICC neurons. Nonlinearities of the neural response which originate in previous auditory structures and are present in the input to the ICC, such as the cochlear rectification or cochlear distortions \citep{McAlpine2004} have been found. Nonlinear response mechanisms might exist that are intrinsic processing properties, originating in the ICC, such as spike generating nonlinearities \citep{MontySpkThres2005}, feedback kernels \citep{GLMWoolley2011} or the suggested neural selectivity for single calls \citep{PortforsOverRepVocAwakeMouse2009}.
We argue that %
if the nonlinearities came from different stations preceding the ICC, then they would show no correlation with the neuron's linear response. % 
If, on the other hand the nonlinearity results from the neuron's own processing, the linear and nonlinear responses likely display similarities. We demonstrated the spectral overlap between the linear and nonlinear responses for the majority of neurons which displayed significant nonlinearities. We showed, respectively for half of the neurons, either same and modulatory temporal tuning for the linear and nonlinear responses. % 
Thus an important part of the nonlinearities do originate in the ICC and further studies of these will reveal more aspects of their sound encoding. \\%
\newpage
Furthermore, no correlation was found between the degree of nonlinearity as assessed by the singular values' magnitude, and the phase-locking or feature-selectivity index. Nonlinear response properties were not linked to the match of the RTF and CRH, or to any particular type of linear STRF. Thus, the found nonlinearities do not seem to be linked to a specific type of neuron but inherent to ICC neurons with very diverse filtering characteristics.\\ 
In contrast to previous studies which used the entire waveform or stimulus envelope, in the present study, the amplitude spectrum for each frequency carrier of the sound stimulus were used to compute the spike-triggered covariance. This allows identifying variations of nonlinear responses across frequencies and to compare the STC's temporal tuning to that of the spectrotemporal receptive field.\\
This unique approach to frequency carrier specific probing of nonlinear response properties can be extended to inter-frequency stimulus correlations to investigate preferences of stimulus correlations from different frequency carriers. %
Earlier studies have indicated nonlinear interactions between different sound frequencies using simple stimuli \citep{EggXFreq2010}.\\ %
Non-classical receptive fields have been suggested to exist in the ICC \citep{WoolleyExtraTuning2011} using spectrotemporal receptive fields. In the present work nonlinear response properties were derived from the neural response using the spike-triggered covariance. %
In contrast to findings in the visual system for which the complex receptive field displaying nonlinear summation properties lies outside the classical receptive field \citep{HubelWiesel1963,AllmanMcGuinness1985a,AllmanMcGuinness1985b}, % 
in this work it was shown, that at least in the spectral dimension these two receptive fields coincide.\\
In the temporal dimension, we found that about 50\(\%\) of the neurons displayed shifted and same tuning pattern and the other half displayed opposite tuning pattern. This indicates that modulatory responses exist in the ICC, as suggested for the visual system \citep{AllmanMcGuinness1985a, AllmanMcGuinness1985b}; natural sensory stimulation of the nonclassical receptive field increases selectivity and decorrelates the responses of neurons, thus increasing sparseness and information transmission \citep{GallantRF2002}. The classical and nonclassical receptive fields might act together as a single processing unit, optimized for natural sound stimuli.\\ 
The presented findings might differ from neural processing in awake animals, as the  recordings were taken from animals that were put under anesthesia which might change inhibitory mechanisms that would affect the nature of the receptive fields and the nonlinearities \citep{SharpeeRFdim2012}. 
Here, we investigated neural responses specifically to the stimulus amplitude spectrum, which is crucial for speech recognition. In a further experiment, it would be interesting to investigate whether neuronal sensitivity to the fine structure is present in the ICC and it might be more pronounced than demonstrated for the auditory cortex since temporal resolution is higher in the ICC than in the cortex \citep{SharpeeRFdim2012}.\\
In order to probe neural encoding of natural sound, our presented method can be employed to study whether single neuron's responsiveness to vocalizations can be directly linked to preferences found with the STRF and within the STC to corresponding stimulus correlations. Whether neural preferences to stimulus correlations that are found in a vocalization correspond to a reliable representation of that vocalization by a single neuron can be investigated via neural discrimination. %
For prediction models that include the linear and nonlinear neural preferences, the spike-triggered method is beneficial, since it provides not only the estimate of the preferred stimulus feature, but also the nonlinear gain function which describes how spiking probability changes as a function of the similarity between the presented stimuli and the optimal stimulus feature \citep{SharpeeReviewHierarchAudio2011}. \\
Using the dynamic moving ripple sound amplitude spectrum for the analysis, it could be shown that dominant nonlinearities were found at the best frequency of the neuron. Previous studies showed a good match between the averaged linear STRF for the whole waveform and the highest significant singular vector of the STC \citep{Lewis2002}. In this work, not only similar temporal courses of the linear STRF and STC were found, but also opposing tuning. Lying within the STRF, the linear receptive field, these nonlinear responses may form a processing unit and have modulatory effects as suggested for neurons in the songbird \citep{WoolleyExtraTuning2011}. \\% 
\newpage
In this work, single neurons were probed for preferences to stimulus envelope correlations and these nonlinear properties were compared to linear spectrotemporal preferences. %
We find that more than half of ICC neurons examined display significant nonlinear response properties generally accompanied by a linear response component. The greatest nonlinearities are found at frequencies near the best frequency that drives the maximal linear response. This suggests a high degree of overlap between the linear and nonlinear receptive field components. In addition, the nonlinear response components have either the same or opposite temporal receptive field pattern as the linear response components. %
This study indicates that the nonlinearities are a general property of many ICC neurons.  Significantly, it supports the notion that non-linear and linear responses are not due to distinct frequency sensitivities (e.g., harmonic) but instead have similar frequency-dependence. This could emerge from ascending auditory pathways known to process different sound features and to converge onto a common frequency-lamina within ICC.
\subsection*{Acknowledgments}
DL was supported by Grant \(\#\) 01GQ0811 of the Federal Ministry of Education and Research within the Bernstein Focus of Neural Technology G\"ottingen. We acknowledge funding support from National Science Foundation (NSF) Award 1355065 to H. L. Read and M. A. Escabí, and NIH-Award 015138 to M. A. Escabí and H. L. Read.\\
%
%
%\subsection*{Author note}
\\
\textbf{Author note} This paper is based on parts of %the Ph.D. thesis of DL 
https://ediss.uni-goettingen.de/handle/11858/00-1735-0000-0022-6026-D %
%by DL, %
which has been published under the Common License agreement. 
\bibliographystyle{jneurosci}
\renewcommand{\bibname}{References}

%\bibliography{lita3}
\end{document}